Spatially resolved observations of Europa's surface with Subaru/IRCS at 1.0–1.8 μm: Upper limits to the abundances of hydrated Cl-bearing salts

Shuya Tan[1,2], Yasuhito Sekine[2,3], Masayuki Kuzuhara[4,5]

Abstract

Recent infrared observations at 1.5–4.0 μm using large ground-based telescopes have suggested that Cl-bearing salts are likely present on Europa's surface as non-ice materials. The chemical compositions of those Cl-bearing salts are key to understanding Europa's ocean chemistry and habitability. Here we report the results of ground-based telescope observations of Europa across two wavelength ranges, 1.0–1.5 and 1.5–1.8 μm, of which the former range includes absorption features owing to some hydrated Cl-bearing salts. We obtained spatially resolved reflectance spectra using the Subaru Telescope/IRCS and the adaptive optics system AO188 with high wavelength resolutions ($\delta\lambda \sim 2$ nm for 1.0–1.5 μm and $\delta\lambda \sim 0.9$ nm for 1.5–1.8 μm) and low noise levels ($1\sigma \sim 1–2 \times 10^{-3}$). We found no clear absorption features at ~1.2 μm caused by hydrated Cl-bearing salts. We estimated that conservative upper limits to the abundances of $MgCl_2 \cdot nH_2O$, $NaClO_4 \cdot 2H_2O$, $Mg(ClO_3)_2 \cdot 6H_2O$, and $Mg(ClO_4)_2 \cdot 6H_2O$ on Europa are 17% (<10% for most) at the $3\sigma$ noise level. These values are lower than the proposed abundance of some hydrated Cl-bearing salts (> ~20%) on Europa based on previous observations. This supports the idea that Cl-bearing salts on Europa are likely anhydrous Na salts of NaCl and/or $NaClO_4$, or hydrated $NaCl \cdot 2H_2O$. The presence of Na salts suggests that $Na^+$ could be the major cation in Europa's ocean, which would be possible if the oceanic pH is circumneutral or alkaline.

[1] Corresponding author shtan@elsi.jp
[2] Earth-Life Science Institute (ELSI), Tokyo institute of Technology, 2-12-1, Ookayama, Meguro, Tokyo, Japan
[3] Institute of Nature and Environmental Technology, Kanazawa University, Kanazawa, Japan
[4] Astrobiology Center, 2-21-1, Oosawa, Mitaka-si, Tokyo, Japan
[5] National Astronomical Observatory of Japan, 2-21-1, Oosawa, Mitaka-si, Tokyo, Japan

## 1. INTRODUCTION

Europa, the second satellite of Jupiter, is thought to have a subsurface ocean and has attracted attention owing to its geological activity and astrobiological potential (e.g., Kivelson et al. 2000). Although geophysical and geological evidence supports the presence of the subsurface ocean, little is known about its water chemistry (e.g., seawater pH, dissolved species, or redox state), knowledge of which is critical for understanding Europa’s habitability. Through water-rock interactions on Europa, cations and anions would be exchanged between the rocky interior and overlying subsurface ocean, thus controlling the ocean’s water chemistry (e.g., Zolotov & Kargel 2009; Tan et al. 2021). The water chemistry of Europa’s ocean has been discussed on the basis of geochemical modeling (Zolotov & Kargel 2009) and laboratory experiments of water-rock interactions (Tan et al. 2021). Depending on the conditions governing the water-rock interactions (e.g., the existence of hydrothermal environments, rock compositions, and $H_2$ fugacity), the resulting water chemistry changes (e.g., Tan et al. 2021). To further constrain the water chemistry and reaction conditions, observational constraints on the major cations and anions in Europa’s ocean are crucial.

Non-ice materials on Europa are considered promising indicators for revealing its water chemistry. These materials have been observed in ridges and chaotic terrains (e.g., McCord et al. 1998; 1999), the latter of which cover equatorial regions and their disrupted ice plates (Doggett et al. 2009). Chaotic terrains may have formed through melting and disruption of pre-existing surface ice by heat from the interior (Greenberg et al. 1999; O’Brien et al. 2002). Non-ice materials on Europa exhibit reddish or yellow colors at visible wavelengths and have relatively low reflectance at near-infrared (NIR) wavelengths, thus weakening the absorption of $H_2O$ ice (e.g., McCord et al. 1998; 1999; Fischer et al. 2015). Based on reflectance data at wavelengths of 0.7–5.2 μm taken by the Near-Infrared Mapping Spectrometer (NIMS) aboard the Galileo spacecraft, sulfate salts and sulfuric acids have been proposed as non-ice materials on Europa (Carlson et al. 1999; Dalton 2007). However, the regions observed by the Galileo spacecraft were spatially limited. In addition, owing to the low S/N ratios and wavelength resolution in Galileo’s NIMS observations (S/N ~ 30, $\delta\lambda$ = 25 nm; Carlson et al. 1992;

Dalton et al. 2012), the interpretation of the observations remains controversial.

Recent spatially resolved observations with large ground-based telescopes (e.g., the Keck II telescope and the Very Large Telescope, VLT) and advanced adaptive optics (AO) systems with high wavelength resolution ($\delta\lambda \sim 0.6$ nm) have suggested that chlorine (Cl)-bearing salts are likely present on geologically young surface areas, including on chaotic terrains on Europa's surface (Brown & Hand 2013; Fischer et al. 2015; 2017; Ligier et al. 2016). It has been suggested that sulfuric acids are distributed mainly in the trailing hemisphere, where exogenic sulfur was implanted from Io (Fischer et al. 2015). The abundances of sulfuric acids are estimated to be as high as 40%–50% in the trailing hemisphere (Fischer et al. 2015; Ligier et al. 2016); whereas, those in the leading hemisphere are 20% or lower (Fischer et al. 2015; Ligier et al. 2016). Brown & Hand (2013) and Fischer et al. (2015) suggested that anhydrous Cl-bearing salts (e.g., NaCl, KCl, and/or small amounts of $MgCl_2$) are likely present in chaotic terrains of the leading hemisphere. On the other hand, Ligier et al. (2016) suggested that Cl-bearing salts composed of hydrated Mg-bearing salts (e.g., $MgCl_2 \cdot nH_2O$) are distributed across chaotic terrains. In their modeling, the relative abundance of Cl-bearing salts was suggested to reach ~20% and ~30% in chaotic terrains of the leading and trailing hemispheres, respectively (Ligier et al. 2016). In addition, a possible color center of NaCl at visible wavelengths was observed in reflectance spectra of chaotic terrains of the leading hemisphere (Trumbo et al. 2019). On the other hand, recent experimental and observational studies suggested the presence of sulfate salts in the trailing hemisphere to explain reflectance spectra obtained at visible and infrared wavelengths (Brown & Hand 2013; Hibbitts et al. 2019; Trumbo et al. 2020).

These prior studies proposed a variety of anhydrous and/or hydrated Cl-bearing salts as candidates for Cl-bearing salts on Europa, including NaCl, $NaCl \cdot 2H_2O$, $NaClO_4$, $NaClO_4 \cdot 2H_2O$, $MgCl_2 \cdot nH_2O$, $Mg(ClO_3)_2 \cdot 6H_2O$, and $Mg(ClO_4)_2 \cdot 6H_2O$ (Fischer et al. 2015; 2017; Ligier et al. 2016; Thomas et al. 2017). Anhydrous Cl-bearing NaCl and $NaClO_4$ salts are insensitive to NIR light at wavelengths of 1.0–2.5 μm; therefore, they exhibit no obvious absorption features in NIR reflectance spectra (Hanley et al. 2014; Figure 1). In contrast, hydrated Cl-bearing salts such as $NaClO_4 \cdot 2H_2O$, $MgCl_2 \cdot nH_2O$, $Mg(ClO_3)_2 \cdot 6H_2O$, $Mg(ClO_4)_2 \cdot 6H_2O$, and $NaCl \cdot 2H_2O$ show characteristic absorption features in NIR spectra owing to O–H bonds in their structures (e.g., absorption peaks at ~1.2, ~1.5, ~1.8, and ~2.0 μm) (Hanley et al. 2014;

Figure 1). The peak positions and shapes of the absorption features of hydrated Cl-bearing salt often resemble those of other hydrated salts (see, e.g., $MgCl_2 \cdot 6H_2O$ and $Mg(ClO_4)_2 \cdot 6H_2O$ in Figure 1). In addition, their absorption positions are located near those of pure $H_2O$ ice (Figure 1). Thus, identification of Cl-bearing salts needs observational data with high wavelength resolution, a wide wavelength range, and a low noise level.

To constrain the composition of Cl-bearing salts on Europa, one problem hampering previous telescope observations was their limitation of the observational wavelength range. Previous studies performed observations at 1.5–1.8 μm (i.e., the *H* band), 1.9–2.5 μm (i.e., the *K* band; Fischer et al. 2015; Ligier et al. 2016), and 3–4 μm (i.e., the *L* band; Fischer et al. 2017). The major absorption centers of hydrated salts mostly overlap with those of $H_2O$ ice and water vapor in the wavelength range used for most previous observations. Ligier et al. (2016) discussed the presence of hydrated Cl-bearing salts based on their absorption tails, which were located near the edges of the wavelength ranges (namely, around 1.5 and 1.8 μm).

To identify and quantify the presence of hydrated Cl-bearing salts on Europa, new observations with high wavelength resolution, a wide wavelength range, and high S/N ratios are required. In the wavelength range 1.15–1.25 μm, some hydrated Cl-bearing salts have characteristic narrow absorption peaks (Figure 1), which are distinct from the broad absorption features of $H_2O$ ice and hydrated sulfate-bearing salts and acids (Hanley et al. 2014). Since telluric atmospheric absorption features are weak in this wavelength range (e.g., Tokunaga et al. 2002), reflectance spectra of Europa at 1.0–1.5 μm (i.e., the *zJH* band) with high S/N ratios and wavelength resolution could provide key constraints on the presence of some hydrated Cl-bearing salts. In addition, an AO system applicable to the *zJH* band is required to spatially resolve Europa. Among large ground-based telescopes, an AO system for the *zJH* band is available on the Subaru 8.2 m telescope at Mauna Kea, Hawaii.

We performed NIR observations using the Subaru telescope across the wavelength range spanning 1.0–1.8 μm (i.e., covering the *zJH* and *H* bands), encompassing the 1.15–1.25 μm wavelength range. In Section 2, we describe the observations, methodology, and data reduction method. In Section 3, we show our results based on the NIR observations of Europa’s surface. In Section 4, we discuss the presence or absence of hydrated Cl-bearing salts on Europa. Using our observational data, combined with reflectance spectra of hydrated Cl-bearing salts, we provide upper limits to the abundances on Europa.

Finally, we summarize our results and suggest implications for non-ice surface materials on Europa in Section 5.

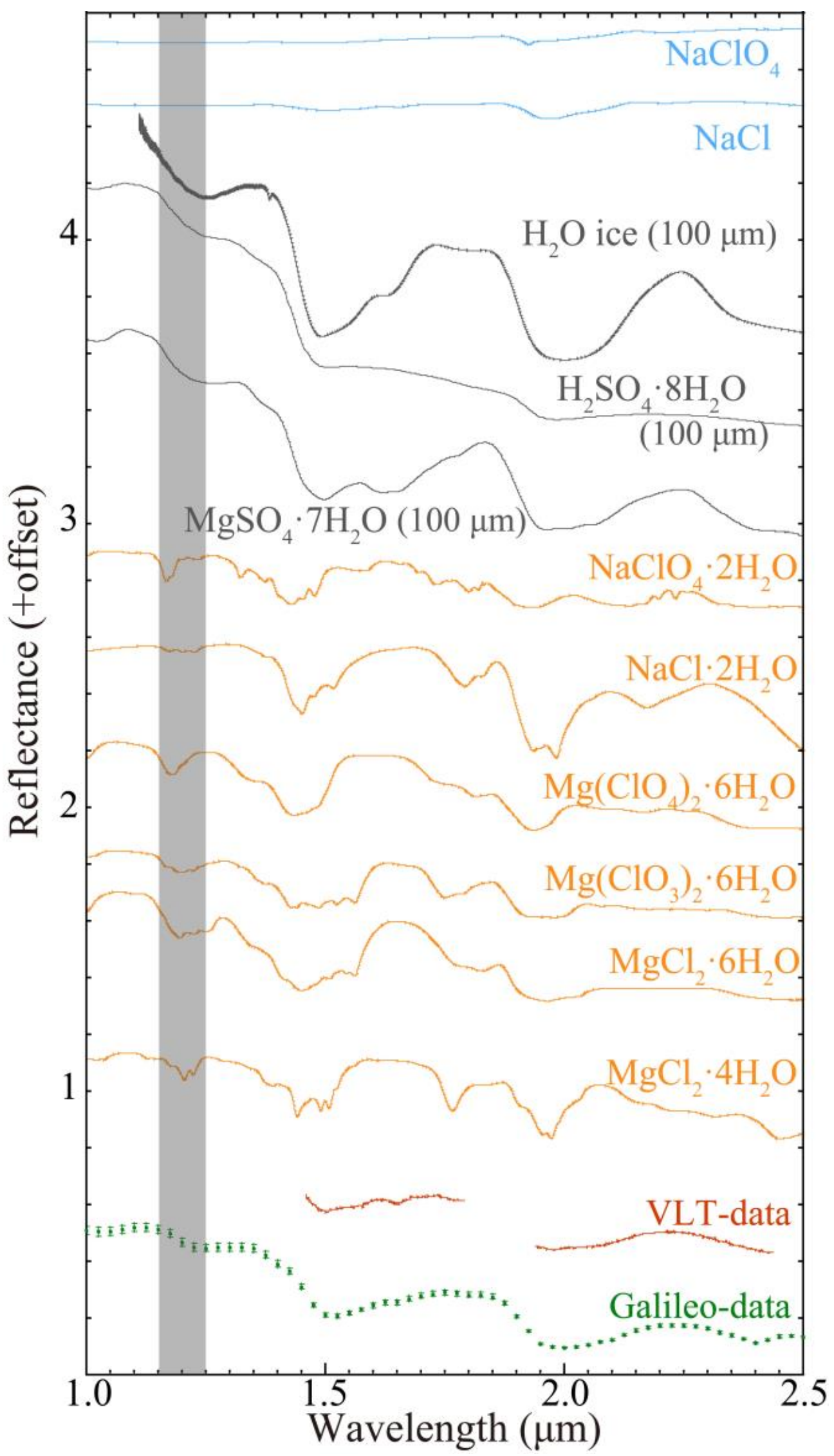

Figure 1. Reflectance spectra of Cl-bearing salts, $H_2O$ ice, $H_2SO_4 \cdot 8H_2O$, $MgSO_4 \cdot 7H_2O$, and Europa's surface. Data of Cl-bearing salts were obtained at 80 K (derived from Hanley et al. 2014), except for $NaCl \cdot 2H_2O$ at 100 K (derived from Fox-Powell et al. 2019). The spectrum of $NaCl \cdot 2H_2O$ is derived from the natural sample possibly containing impurities. VLT and Galileo/NIMS spectra were obtained at coordinates 195°W, 0°N and 130°–260°W, 35°S–35°N and taken from Ligier et al. (2016) and McCord et al. (1999), respectively. Reflectance spectra of $H_2O$ ice, $H_2SO_4 \cdot 8H_2O$, and $MgSO_4 \cdot 7H_2O$ were calculated using the Hapke model (Hapke 1981; 1993; 2002) with the optical constants for amorphous $H_2O$ ice at 70–130 K (Mastrapa et al. 2008), $H_2SO_4 \cdot 8H_2O$ at ~77 K (Carlson et al. 2005), and $MgSO_4 \cdot 7H_2O$ at 120 K (Dalton & Pitman 2012). Reflectance spectra are shown with positive offsets for clarification. Gray bars represent the wavelengths

corresponding to possible absorptions of hydrated Cl-bearing salts (1.15–1.25 μm).

## 2. OBSERVATIONS AND DATA REDUCTION

### 2.1. Observations using Subaru/IRCS

The observations were performed using the Infrared Camera and Spectrograph (IRCS; Tokunaga et al. 1998; Kobayashi et al. 2000) and AO system (AO188) (Hayano et al. 2008; 2010) installed on the Subaru Telescope. The spectrograph was set to use the 7.04" × 0.30" slit in the grism mode covering the *zJH* band (0.95–1.5 μm, $\delta\lambda \sim 2$ nm) and the *H* band (1.49–1.83 μm, $\delta\lambda \sim 0.9$ nm), characterized by respective wavelength resolutions of ~20 and ~30 times higher than that of Galileo's NIMS. The observations yielded two-dimensional spectral images. The images were acquired in 1024 × 1024 pixels format with ALADDIN III InSb arrays (Kobayashi et al. 2000). The pixel size is 0.05277" ± 0.00004". The spatial resolution of the settings at wavelengths of 1–2 μm was reported as 0.05–0.1" (Minowa et al. 2010).

Table 1 summarizes our observations. The leading hemisphere of Europa was observed on 2019 May 17 from 11:58 to 13:41 Universal Time (UT); whereas, the trailing hemisphere was observed on 2019 May 18 from 11:21 to 13:06 UT (Table 1). We obtained the precise viewing geometry and ephemerides from the NASA JPL Horizons Web Interface interactive system (http://ssd.jpl.nasa.gov/). The slit was aligned East–West as applied to Europa, having adopted the north pole position angle based on NASA JPL Horizons (Figure 2).

Before and/or after the observations of Europa, we observed solar-type (G2V-type) stars using an ABBA nodding pattern (Table 1). Specifically, we observed HD 154805 in the *zJH* band and HD 160257 in the *H* band, under similar airmasses as those of our Europa observations. The data of the stars were used for telluric atmospheric absorption corrections and to determine the instrumental characteristics (labeled 'Telluric calibrator' in Table 1). The data of the stars were also used for calculations of Europa's reflectance spectra (see Section 2.3). In the present study, B-type stars (HD 154204, HD 165805, and HD 157546) were observed under similar airmasses to check the validity of the data reduction procedures (see Section 2.3).

Each observation of Europa consisted of 4–8 pointings, where the targets were dithered between opposite ends along the slit in an ABBA pattern. In order to increase the S/N ratio, we averaged and stacked all spectra to generate a single spectrum.

To adjust the parameters of the AO system, the integration times, and the slit alignments for all targets, we also observed our targets in the IRCS imaging mode in the *H* band (at a central wavelength of 1.63 ± 0.15 μm) and the *K* band (at a central wavelength of 2.20 ± 0.17 μm) immediately before using the grism mode. The images were captured with and without the slit setting. The images were also used to analyze the areas observed on Europa as well as the uncertainties in the data of the G2V stars (see Sections 2.2 and 2.3 for our analysis).

We performed general spectral image reduction (namely sky subtraction, flat-fielding, corrections for bad pixels, and corrections for cosmic-ray events) and extraction of one-dimensional spectra from the two-dimensional spectral images using the IRAF package (Tody 1986; 1993). Since Europa is a spatially extended object, unlike a point source target of a star, Europa's spectra were extracted from two-dimensional spectra spread across a width of 40 pixels in the observed images: see Figure 3(a). Wavelength calibration was performed using IRAF with the aid of the emission-line spectrum of an Ar lamp and comparison images. The images observed in imaging mode were also corrected by application of general spectral image reduction, again using the IRAF package. For flat-fielding, we used archival flat-field images obtained in the IRCS imaging mode on 2019 May 19 in the SMOKA (Baba et al. 2012) with the same settings as those used for our observations, since our observations did not initially include flat-field images in imaging mode.

## 2.2. Analysis of the areas observed on Europa

The slit positions on Europa were determined on a pixel scale based on the images captured in the IRCS imaging mode (Figure 4). We calculated the slit width of 0.30" as spanning ~5 pixels based on the detector's resolution of 0.05277" pixel$^{-1}$. Using the images of Europa obtained with this slit setting (Figure 4(a)), the slit position (in units of pixels) was applied to the disk of Europa (Figure 4(b)). In addition, the center positions of Europa's disk in the images were determined based on Europa's signal strength. Since the apparent diameter of Europa was ~0.99" at the time of the observations, the size of Europa's disk was calculated as spanning 19 pixels given the detector's resolution. This way, the precise areas observed on Europa's disk were determined.

Based on the pixel scale and the spatial resolution of the setting at wavelengths of 1–2 μm (see Section 2.1), we separated the two-dimensional spectra of Europa into sections with a width of 1.6 pixels along the direction of the slit: see Figure 3(b). This width of 1.6 pixels was calculated to correspond to ~0.087", comparable to the spatial resolution. We performed extraction of one-dimensional spectra for each section from the two-dimensional spectra. The sections correspond to 300–400 km areas from East to West on Europa’s surface. Its surface was also separated into ~1000 km areas from North to South by the 0.30" slit width. The locations of each area on Europa’s surface were determined in units of pixels, based on those slit positions that overlapped with Europa. Next, these locations were converted into coordinates on Europa’s surface using the appropriate viewing geometry and ephemerides (Figure 2).

Spectra of Europa’s surface were obtained for seven separate areas during each night (Figure 5). As part of the observations of 2019 May 17, spectra were obtained for the southwestern regions of the leading hemisphere (Figure 5). Chaotic terrains were included in the observed area around ~150°–100°W and ~20°–40°S in the leading hemisphere (Figure 5), where previous observations suggested the presence of Cl-bearing salts (Fischer et al. 2015; Ligier et al. 2016). Reflectance spectra of the southern regions of the trailing hemisphere were observed on 2019 May 18. The presence of high abundances of sulfuric acids has been suggested near the center of the trailing hemisphere (Fischer et al. 2015; Ligier et al. 2016). In addition to the observational target areas discussed above, we also acquired data for the northern and equatorial regions on Europa’s surface. Since the slopes of the reflectance spectra in the northern and equatorial regions become steeper than those in the Galileo/NIMS data (see Appendix C for details), we used a part of the data from the northern and equatorial regions for detailed analysis. We only used the data of two regions (160 ± 9°W, 3 ± 15°N and 256 ± 5°W, 15 ± 14°N) in the northern and equatorial regions for the estimations for hydrated Cl-bearing salts (Figure 5; see below in Section 4). However, there are no clear absorption features owing to hydrated Cl-bearing salts in the spectra from either the northern or equatorial regions (Figure C2), nor are there in the regions analyzed (see below). Thus, our conclusion of an apparent absence of some hydrated Cl-bearing salts (e.g., $MgCl_2 \cdot nH_2O$) on Europa does not change significantly. Our observational areas do not include the large chaotic terrain, Tara Regio (~85°W, ~0°N), where a possible color center of NaCl at visible wavelengths was observed (Trumbo et al. 2019).

Table 1. Summary of observations of Europa and stars on 2019 May 17 and 18 using the Subaru Telescope.

| Time (UT) | Target | Integration Time (s) | Airmass | Europa's location | | Spectral Type | Spectral Band | Comments |
|---|---|---|---|---|---|---|---|---|
| | | | | Longitude | Latitude | | | |
| 2019 May 17 | | | | | | | | |
| 10:36 | HD 154204 | 0.5 × 40 | 1.390 | | | B7V | *zJH* | |
| 10:53 | HD 165805 | 0.5 × 40 | 1.453 | | | B3V | *zJH* | |
| 11:22 | HD 154805 | 2 × 48 | 1.382 | | | G2V | *zJH* | Telluric calibrator |
| 11:59 | Europa | 0.41 × 40 | 1.355 | 157°W | 4°N | | *zJH* | |
| 12:07 | Europa | 0.41 × 40 | 1.354 | 157°W | 23°N | | *zJH* | |
| 12:13 | Europa | 0.41 × 80 | 1.354 | 155°W | 58°N | | *zJH* | |
| 12:19 | Europa | 0.41 × 40 | 1.355 | 160°W | 35°S | | *zJH* | |
| 12:24 | Europa | 0.41 × 120 | 1.357 | 161°W | 54°S | | *zJH* | |
| 12:37 | HD 154204 | 1 × 80 | 1.347 | | | B7V | *zJH* | |
| 12:55 | HD 154805 | 2 × 96 | 1.432 | | | G2V | *zJH* | |
| 13:19 | Europa | 2 × 48 | 1.431 | 162°W | 1°N | | *H* | |
| 13:23 | Europa | 2 × 48 | 1.443 | 162°W | 22°N | | *H* | |
| 13:29 | Europa | 3 × 112 | 1.459 | 161°W | 55°N | | *H* | |
| 13:35 | Europa | 2 × 48 | 1.476 | 165°W | 35°S | | *H* | |
| 13:40 | Europa | 3 × 112 | 1.492 | 167°W | 55°S | | *H* | |
| 13:56 | HD 160257 | 5 × 192 | 1.511 | | | G2V | *H* | Telluric calibrator |
| 14:09 | HD 157546 | 1 × 80 | 1.510 | | | B8V | *H* | |

Table 1. (continued)

| Time (UT) | Target | Integration Time (s) | Airmass | Europa's location | | Spectral Type | Spectral Band | Comments |
|---|---|---|---|---|---|---|---|---|
| | | | | Longitude | Latitude | | | |
| 2019 May 18 | | | | | | | | |
| 10:47 | HD 154204 | 0.5 × 80 | 1.360 | | | B7V | *zJH* | |
| 11:03 | HD 154805 | 2 × 48 | 1.400 | | | G2V | *zJH* | |
| 11:22 | Europa | 0.41 × 80 | 1.381 | 256°W | 11°S | | *zJH* | |
| 11:29 | Europa | 0.41 × 80 | 1.373 | 256°W | 15°N | | *zJH* | |
| 11:36 | Europa | 0.5 × 160 | 1.365 | 254°W | 57°N | | *zJH* | |
| 11:45 | Europa | 0.41 × 80 | 1.359 | 259°W | 43°S | | *zJH* | |
| 11:55 | Europa | 1.2 × 160 | 1.355 | 263°W | 69°S | | *zJH* | |
| 12:12 | HD 154805 | 2 × 48 | 1.378 | | | G2V | *zJH* | Telluric calibrator |
| 12:26 | HD 154204 | 0.5 × 40 | 1.337 | | | B7V | *zJH* | |
| 12:30 | HD 154204 | 2 × 96 | 1.343 | | | B7V | *H* | |
| 12:44 | HD 160257 | 3 × 112 | 1.374 | | | G2V | *H* | Telluric calibrator |
| 12:57 | Europa | 2 × 144 | 1.397 | 263°W | 10°S | | *H* | |
| 13:04 | Europa | 2 × 144 | 1.411 | 264°W | 40°S | | *H* | |

Note. The apparent planetographic longitude and latitude of Europa on the center of the slit were provided by JPL Horizons (http://ssd.jpl.nasa.gov/). The "telluric calibrator" in the Comments column refers to data used for correctness of the telluric atmospheric absorption features and for calculations of the reflectance spectra (see the main text for details).

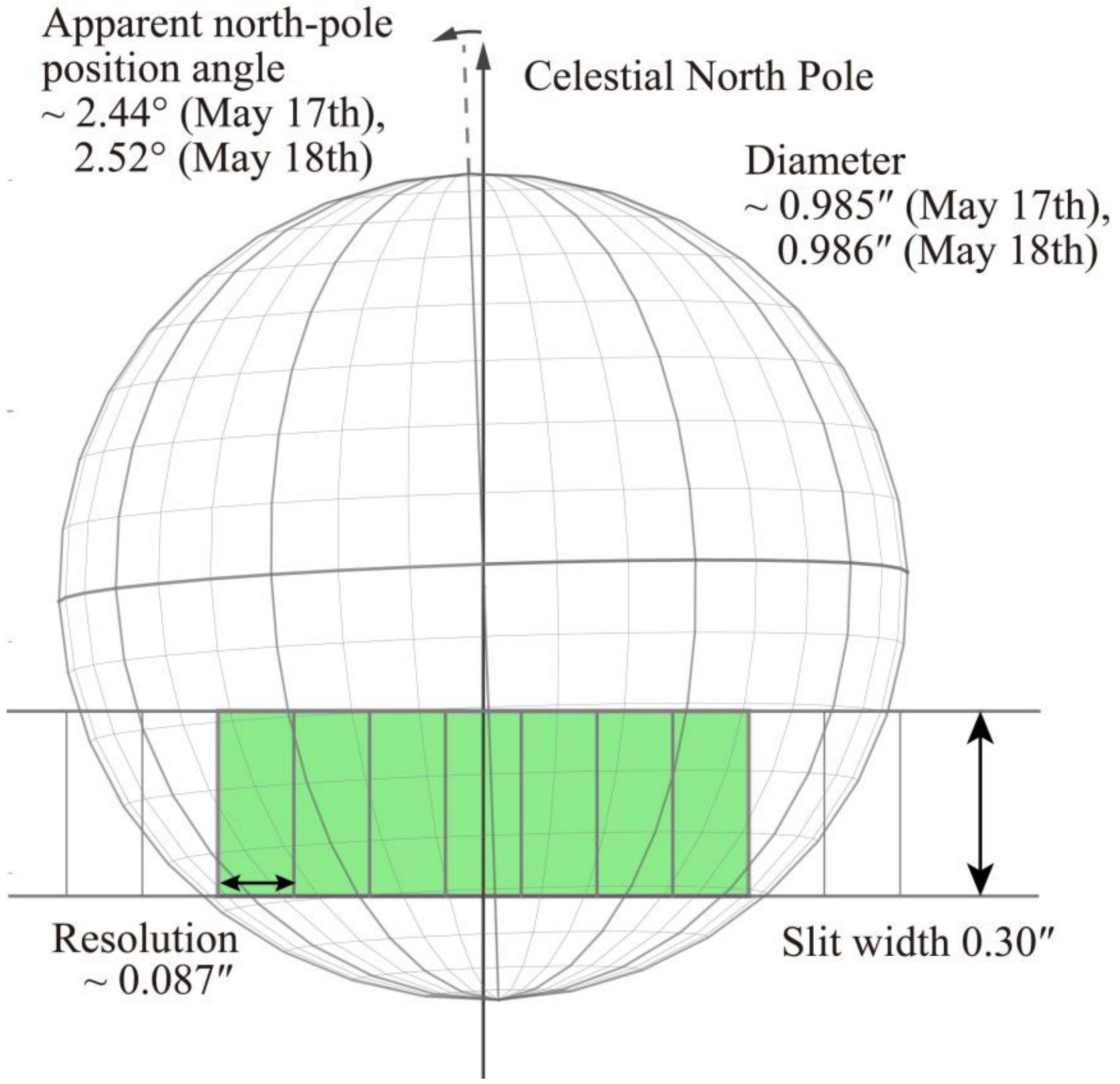

Figure 2. Illustration of the alignment of the slit for our observations of Europa. The slit was aligned in the east–west direction as applied to Europa. On Europa’s surface, the images were separated by ~0.087" in the slit-length direction. Apparent north-pole position angles are the angles of Europa’s north pole with respect to the direction of the celestial north pole. The angular diameters and apparent north-pole position angles were provided by NASA JPL Horizons (http://ssd.jpl.nasa.gov/).

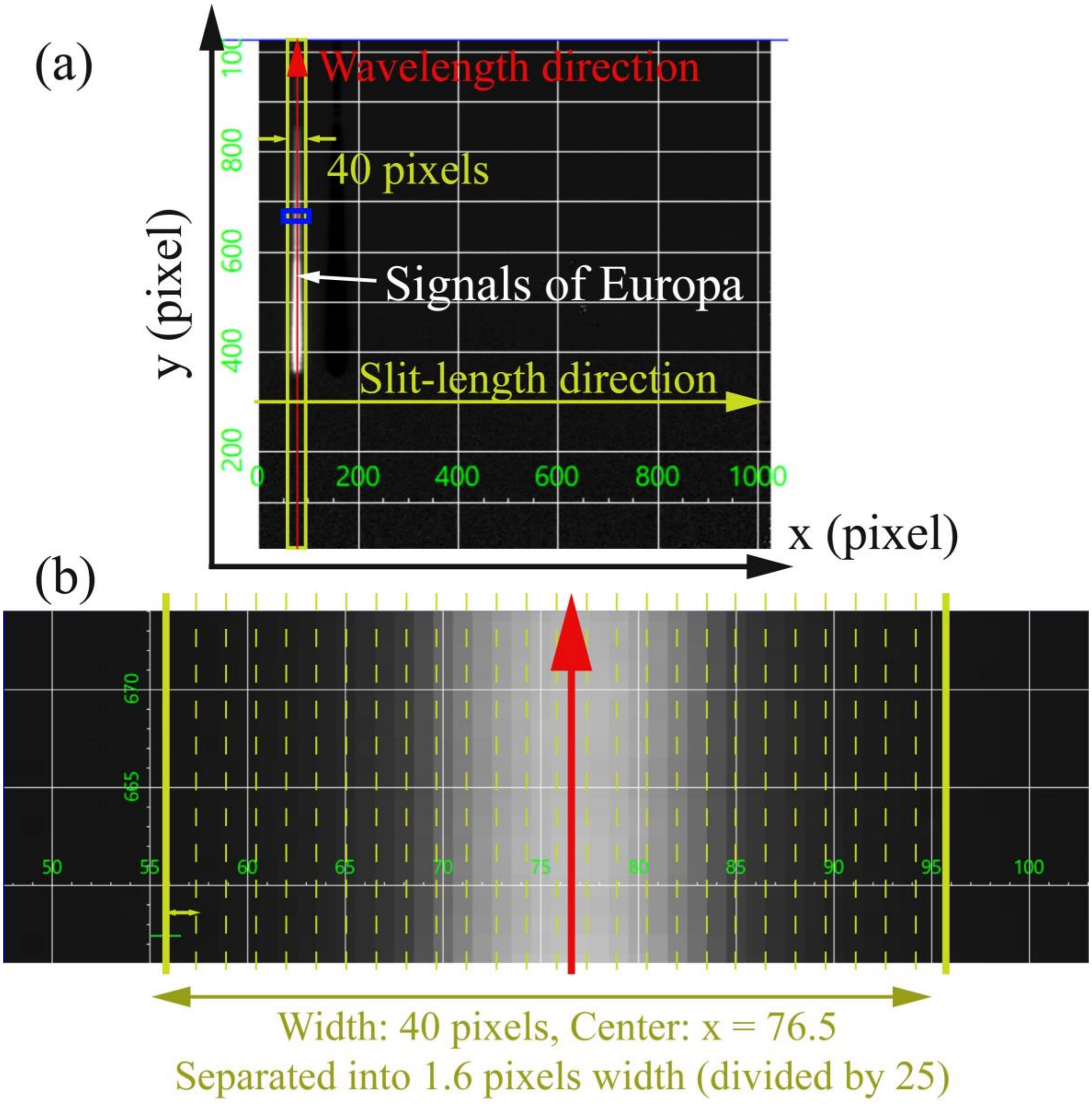

Figure 3. Signal strength of Europa in the two-dimensional spectral images. (a) Europa's signal strength extends in the wavelength direction. The flux distribution was extracted to produce two-dimensional spectra with a width of 40 pixels (see the narrow yellow rectangle). (b) Close-up of the two-dimensional spectra of Europa, corresponding to the range of the small blue rectangle in panel (a). Two-dimensional, 1.6 pixel-wide spectra were extracted in the direction of the slit length. Yellow dashed lines enclose sections with a width of 1.6 pixels each, corresponding to the spatial resolution of the Subaru AO system (~0.087").

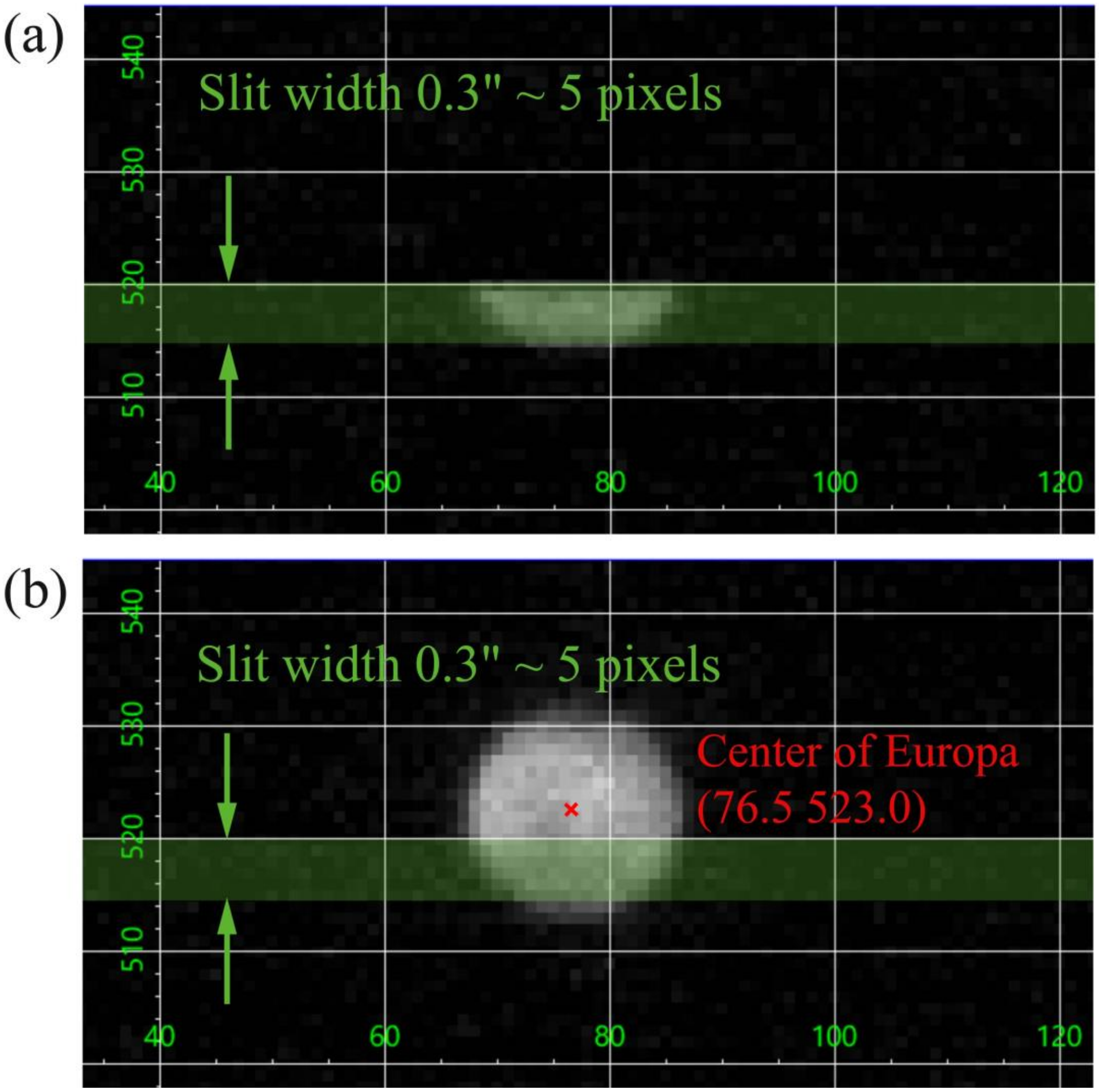

Figure 4. Example of the slit position and location of Europa on the image as captured by the IRCS imaging mode on 2019 May 18. (a) Image with slit setting. The overlying slit was cut out of the disk of Europa. (b) Image without slit setting. The estimated slit position in panel (a) was applied to the green horizontal strip.

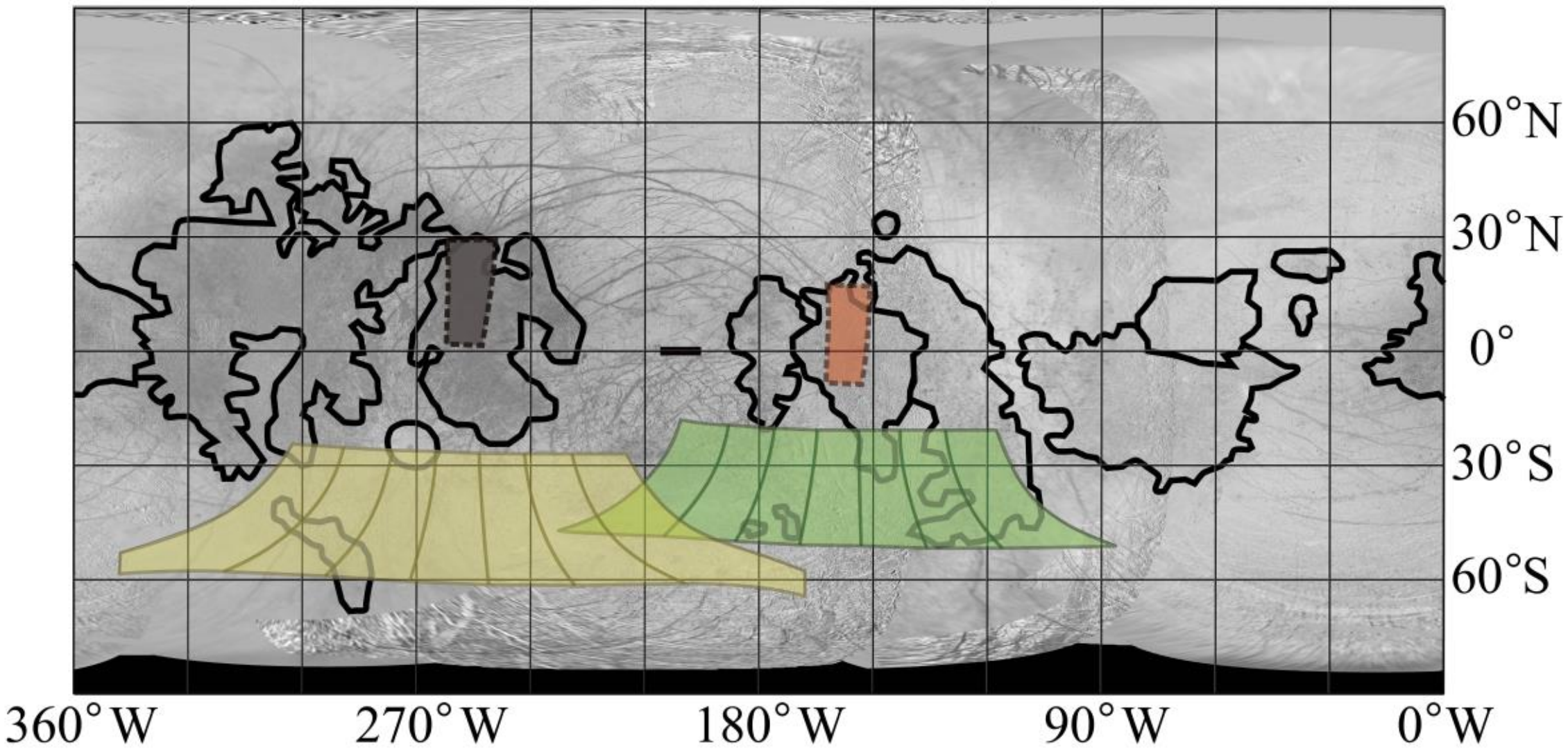

Figure 5. Observed regions on Europa’s surface map (image credit: USGS Map-a-planet). The green and yellow areas were observed on 2019 May 17 and 18, respectively. The green region covers the southeastern region of the trailing hemisphere (210°–180°W, 20°–50°S) and the southwestern region of the leading hemisphere (180–110°W, 20–50°S). The yellow area covers the southern region of the trailing hemisphere (330°–190°W, 30°–60°S). The gray area covers a part of the northern regions of the trailing hemisphere (256 ± 5°W, 15 ± 14°N). The red area covers a part of the equatorial regions of the leading hemisphere (160 ± 9°W, 3 ± 15°N). The spectral data of two areas are used for the estimations for hydrated Cl-bearing salts (see detailed in Section 4). The regions outlined by bold black lines enclose chaotic terrains (Doggett et al. 2009).

### 2.3. Calculations of Europa’s reflectance spectra

Using our spectra of both Europa and the standard G2V stars, a reflectance spectrum, $R(\lambda)$, of a given area on Europa was calculated using the following equation (e.g., Ligier et al. 2016):

$$\boldsymbol{R(\lambda)} = \frac{\textbf{Europa spectra }(\boldsymbol{\lambda})}{\textbf{G2V star spectra }(\boldsymbol{\lambda})} \times \frac{\boldsymbol{T}_{\textbf{G2V star}}}{\boldsymbol{T}_{\textbf{Europa}}} \times \frac{\textbf{1}}{\boldsymbol{\Omega}_{\textbf{area}} \times \textbf{10}^{(\boldsymbol{M}_{\textbf{G2V star}} - \boldsymbol{M}_{\textbf{Sun at Europa}})/\textbf{2.5}}} \times \frac{\textbf{1}}{\textbf{cos }(\boldsymbol{i})}, \quad (1)$$

where $T_{\mathrm{G2V\ star}}$ and $T_{\mathrm{Europa}}$ are the total integration times of Europa and the G2V stars, respectively (see Table 1); $\Omega_{\mathrm{area}}$ is the solid angle of a given distinct area on Europa’s surface (= $5.64 \times 10^{-13}$ sr); $M_{\mathrm{Sun\ at\ Europa}}$ is the apparent magnitude of the Sun seen from Europa (-23.16 mag); $M_{\mathrm{G2V\ star}}$ is the apparent magnitude of the G2V stars (8.75 mag for HD 154805 and 8.58 mag for HD 160257; Hog et al. 2000); and cos($i$) is the photometric correction factor (Ligier et al. 2016). Photometric corrections were applied to obtain geometrically corrected spectra for each area on the surface. Since we approximated Europa's surface by a Lambertian surface, the photometric correction factor can be calculated as follows (Ligier et al. 2016):

$$\cos(i) = \sqrt{1 - \left(\frac{SSP}{r}\right)^2}, \qquad (2)$$

where $r$ is Europa's radius (in pixels) and $SSP$ is the distance (in pixels) between a given target area and the subsolar point on Europa's surface. The planetographic longitude and latitude of the subsolar point at the observation times were provided by NASA JPL Horizons. The angle $i$ corresponds approximately to the phase angle. When $i > 60°$, the corrected reflectance becomes larger than unity based on equation (2). This is because the approximation of a Lambertian surface is invalid near the apparent edges of Europa. Thus, areas with $i > 60°$ were removed from the analysis. Since each area is spreading on Europa's surface (Figure 2), the range encompassed by the angles $i$ caused uncertainties in the absolute values of the calculated reflectance. For instance, $SSP$ ranges from 6 to 8 pixels in the area near the center of the leading hemisphere (111 ± 25°W, 38 ± 16°S), which corresponds to cos($i$) = 0.4–0.7, assuming a radius for Europa of 9 pixels. In the center of the area, cos($i$) = 0.6. Thus, the reflectance of the area is affected by an uncertainty of ~16%–37%. Differences in cos($i$) caused uncertainties in the reflectance of ~10%–37% and ~12%–50% on 2019 May 17 and 18, respectively.

To evaluate the error in the absolute value of the reflectance, we measured the fluxes of the two G2V standard stars on the images captured in IRCS imaging mode without the slit setting. The ratios of the fluxes of the two stars in the same band provide the uncertainty in the absolute value of the reflectance. The ratios of the fluxes of HD 154805 with respect to those of HD 160257 were calculated as 1.14 on 2019 May 17 and 0.92 on 2019 May 18. These ratios took into consideration the difference in brightness of the two stars. In addition to an uncertainty in cos($i$), the absolute values of the reflectance may contain uncertainties of ~10%–54% on 2019 May 17 and ~12%–63% on 2019 May 18.

The present study sets the width of each observed area on Europa's surface to 1.6 pixels, corresponding to the spatial resolution of the Subaru Telescope equipped with AO188. Fluxes from surrounding areas would also be counted as part of the target area since Europa is spatially extended. FWHMs of the point-spread function (PSF) of the fluxes were calculated at about 3 pixels from the images of the G2V standard stars. Assuming Gaussian light distributions with a FWHM of 3 pixels, the fluxes from areas outside 1.6 and 3.2 pixels in a given target area could add up to~23% and ~3% of the total flux in the area, respectively. In addition, flux contributions from surrounding areas would depend on regions on Europa’s disk. This might result in upper or lower estimates of absolute reflectance. For instance, the absolute reflectance would change in a reduction of ~5% near the edge of the disk (e.g., 111 ± 25°W, 38 ± 16°S) and an increase of ~4% on the center of the disk (e.g., 164 ± 11°W, 36 ± 16°S). These variations would not affect our conclusion since they are smaller than the uncertainties due to other factors (e.g., cos(*i*)).

## 3. RESULTS OF THE OBSERVATIONS

### 3.1. Reflectance spectra of Europa’s surface at 1.0–1.8 μm

Figure 6 shows the calculated reflectance spectra of Europa’s surface observed on May 17 and 18 at wavelengths of 1.0–1.5 μm and 1.5–1.8 μm using equation (1) (the original data are shown in Figures A1 and A2 in Appendix A). Due to the uncertainty in the location of Europa’s surface combined with calibration issues (Section 2.3), the reflectance spectra can be shifted upward or downward within the gray range outlined in Figure 6. However, the features of the reflectance spectra do not change.

Although the two reflectance spectra in the *zJH* and *H* bands at a given location are discontinuous at 1.5 μm, the spectra overlap at 1.5 μm within the uncertainties in terms of their absolute values (Figure 6). Considering that the reflectance spectra can be shifted parallel to each other within the uncertainties, we obtained the associated reflectance spectra at wavelengths of 1.0–1.8 μm by connecting the respective reflectance spectra in the *zJH* and *H* bands at 1.5 μm. The spectra were connected by shifting both reflectance spectra, so that the absolute value of the reflectance at 1.5 μm became the mean value of both reflectance spectra.

Figure 7 shows the resulting reflectance spectra of Europa’s surface in the wavelength range 1.0–1.8 μm. The wavelength resolutions, δλ, are ~2.6 nm at 1.0–1.5 μm and ~2.0 nm at 1.5–1.8 μm. These values were calculated based on the FWHMs of the water-vapor absorption features. Figure 7(a) shows the results for the southeastern regions of the trailing hemisphere and the southwestern regions of the leading hemisphere. The absolute reflectance in the wavelength range 1.5–1.8 μm is 0.2–0.3 in the observed regions, which is consistent with the equivalent values observed using other large telescopes (Fischer et al. 2015; Ligier et al. 2016): see Figure 7(a). This figure indicates that the spectra have broad absorption features around 1.3, 1.5, and 1.65 μm caused by $H_2O$-ice absorption (see also Figure 1; Ockman 1958; Carlson et al. 2009). Clear absorption features associated with crystalline $H_2O$-ice are also seen at 1.31 μm (e.g., Grundy & Schmidt 1998). Figure 7(a) shows that none of the spectra have obvious absorption features, other than those caused by $H_2O$ ice. In particular, no clear absorption features are found at a wavelength of ~1.2 μm, where some hydrated Cl-bearing salts exhibit their narrow absorption features (Figure 1). This suggests low abundances of some hydrated Cl-bearing salts on the surface, including $NaClO_4 \cdot 2H_2O$, $MgCl_2 \cdot nH_2O$, $Mg(ClO_3)_2 \cdot 6H_2O$, and $Mg(ClO_4)_2 \cdot 6H_2O$ (see Section 4 for a quantitative discussion as to the upper limit to the abundance).

Figure 7(a) shows that the slopes at 1.5–1.8 μm become steeper in regions closer to the center of the leading hemisphere (90°W, 0°N). Previous studies suggested a high abundance of sulfuric acid in the trailing hemisphere (Fischer et al. 2015; Ligier et al. 2016) due to the implantation of S originating from the inner satellite Io and subsequent chemical reactions on Europa (Carlson et al. 2002; 2005; Dalton et al. 2013). Given that the reflectance spectra of sulfuric acid have a blue slope in the wavelength range 1.0–2.5 μm (Figure 1), the observed trend of steeper slopes in regions closer to the center of the leading hemisphere would be due to the low abundance of sulfuric acids. No clear absorption features at ~1.2 μm are found in spectra covering chaotic terrains (111 ± 25°W, 38 ± 16°S) or other regions (e.g., 178 ± 14°W, 36 ± 16°S), suggesting that the abundances of some hydrated Cl-bearing salts (e.g., $MgCl_2 \cdot nH_2O$) are also low in chaotic terrains.

Figure 7(b) shows reflectance spectra for the southern regions of the trailing hemisphere. Similarly to Figure 7(a), this figure reveals no clear absorption features, including absorption at 1.2 μm, except for absorption owing to $H_2O$ ice. This figure also shows that the slopes at 1.5–1.8 μm become gentle in regions close to the center of the trailing hemisphere (270°W, 0°N), which is consistent with the high abundances of sulfuric acids near the center of the trailing hemisphere (Fischer et al. 2015; Ligier et al. 2016).

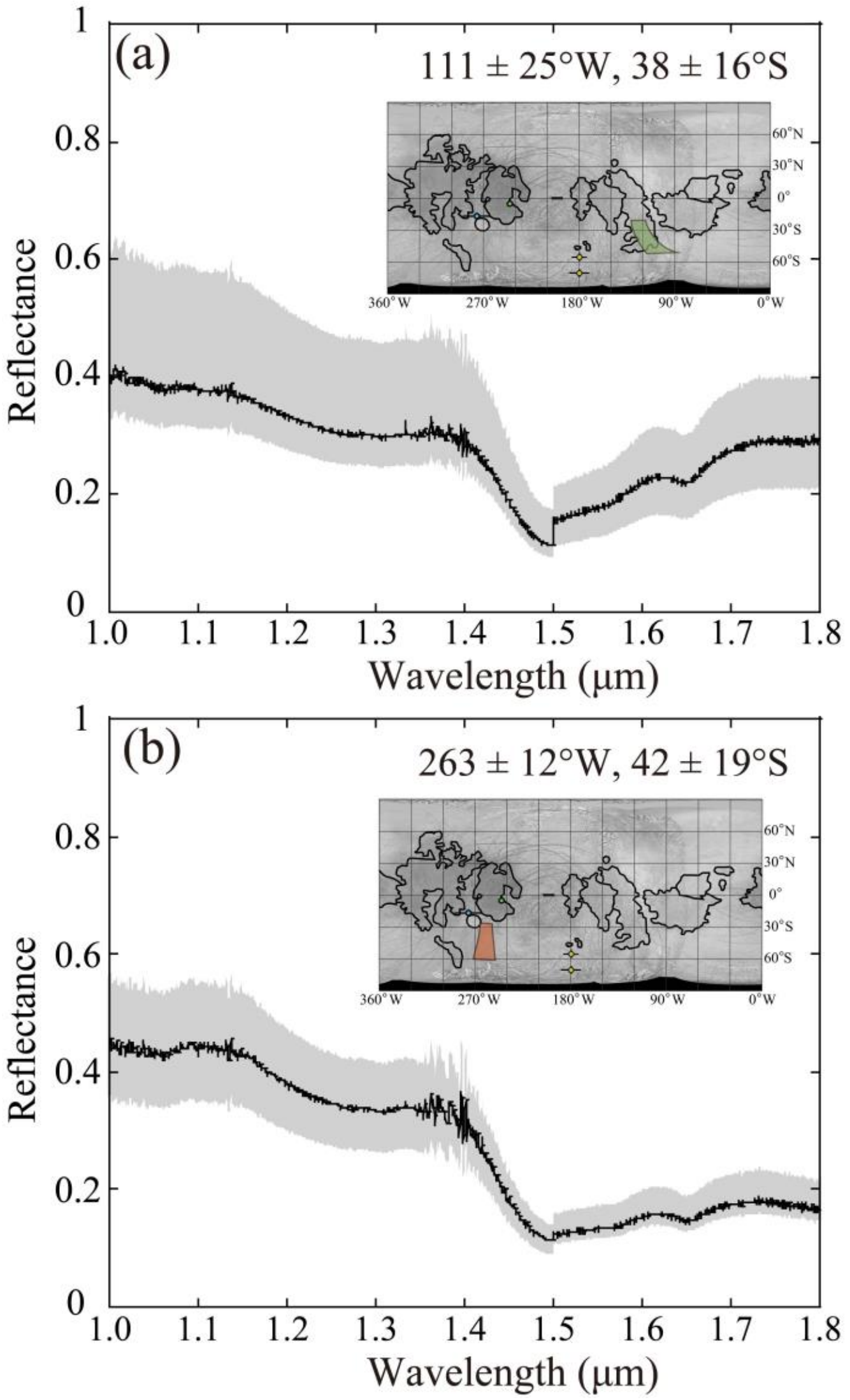

Figure 6. Typical reflectance spectra of Europa’s surface. Each curve shows spectra of one area (shown as colored regions in the top-right insets). (a) Data for the southern regions in the leading hemisphere observed on 2019 May 17. (b) Data for the southern regions in the trailing hemisphere observed on 2019 May 18. The black curves were obtained using equation (1). Gray areas represent the uncertainty in the absolute value of the reflectance due to the uncertainties in the photometric correction factors, cos($i$), and differences in the observing conditions of the G2V stars (see Section 2.3). The reflectance spectra can be shifted within the gray areas; however, the spectral features do not change.

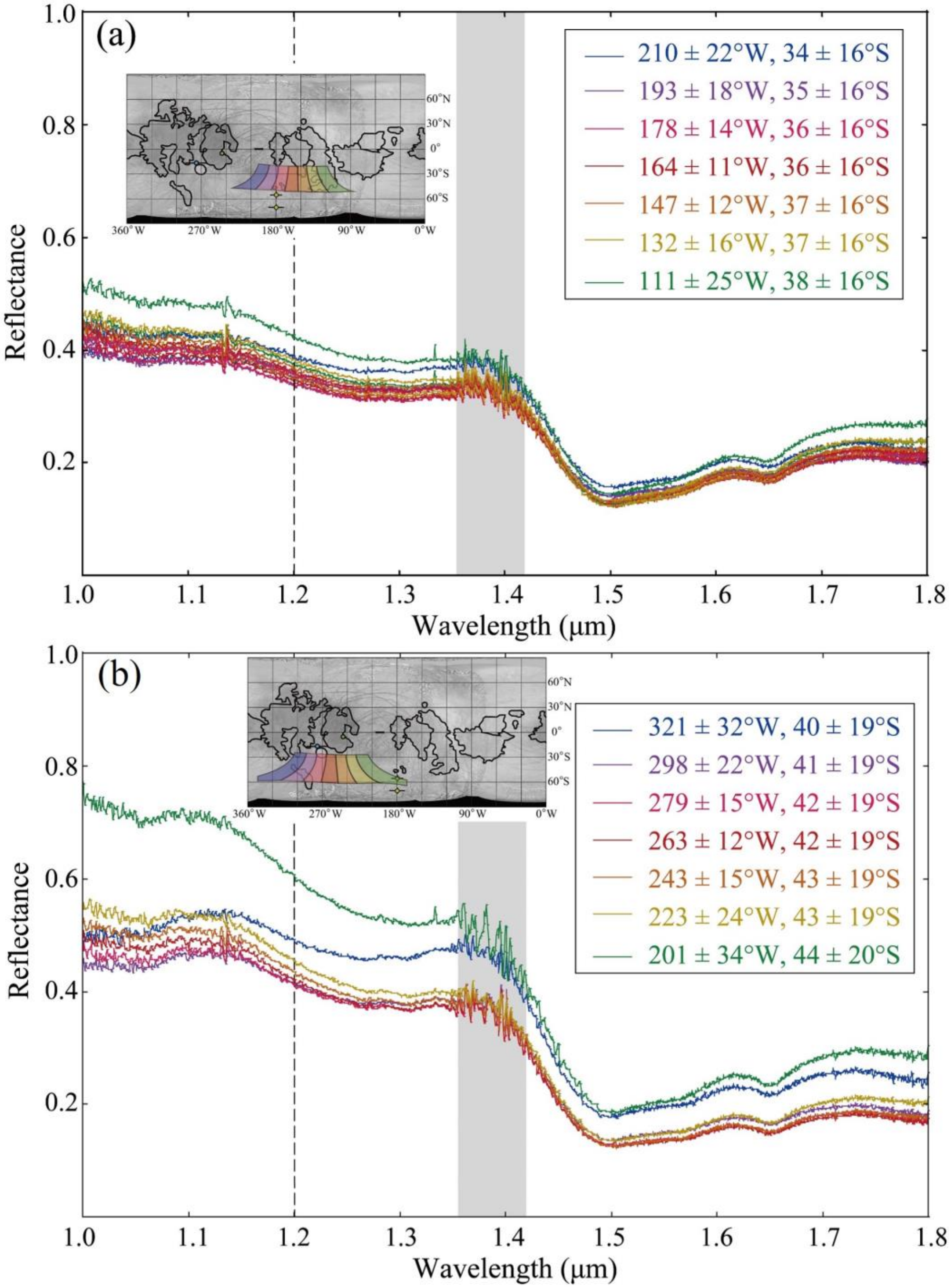

Figure 7. Connected reflectance spectra of Europa's surface at wavelengths of 1.0–1.8 μm. The observed area on Europa is shown in the top-left inset. (a) Data for the southern regions in the leading hemisphere observed on 2019 May 17. (b) Data for the southern regions in the trailing hemisphere observed on 2019 May 18. Gray bars represent the wavelengths corresponding to telluric atmospheric absorption (1.35–1.42 μm). Vertical dashed lines indicate a wavelength of 1.2 μm. In the reflectance spectra, we observe no significant absorption features around 1.2 μm, where hydrated Cl-bearing salts exhibit their characteristic absorption features.

## 4. DISCUSSION

Our observational results in Figure 7 show no clear absorption features at ~1.2 μm in all of the reflectance spectra covering Europa's surface. To further assess the presence or absence of the hydrated Cl-bearing salts, the present study fits the library spectra of $H_2O$ and/or $H_2SO_4$ ice to the observed spectra at the wavelengths excluding 1.150–1.255 μm. Figure 8(a) shows the observed reflectance spectra at 111 ± 25°W, 38 ± 16°S at 1.100–1.300 μm, as an example. Figure 8(b) shows the differences between the reflectance spectra and the fitted spectra of amorphous $H_2O$ ice, or crystalline $H_2O$ ice plus $H_2SO_4$ ice, obtained using the least-squares method at wavelengths of 1.100–1.130 and 1.255–1.300 μm, where there are little absorptions due to hydrated Cl-bearing salts. The reflectance spectra at 1.13–1.15 μm were not used for fitting since a high level of noises is present around 1.13–1.15 μm due to absorption by water vapor in the telluric atmosphere. The spectra of $H_2O$ ice were calculated using the Hapke model (Hapke 1981; 1993; 2002) with the optical constants for amorphous $H_2O$ ice at 70–130 K (Mastrapa et al. 2008), crystalline $H_2O$ ice at 120 K (Grundy & Schmidt 1998), and $H_2SO_4$ ice at 77 K (Carlson et al. 2005). The optical constants of $H_2O$ ice were supplemented with optical constants from Warren & Brandt (2008) to cover the wavelength range of 1.100–1.130 μm. The differences between the reflectance spectra and the fitted spectra (Figure 8(b)) shows that there are no clear absorptions in wavelengths of 1.150–1.255 μm, comparable to those due to hydrated Cl-bearing salts. The results of no significant absorption features in 1.150–1.255 μm are generally common for the observed regions. To remove the continuum across the wavelength range, we also attempted to achieve fits of the observed spectra by polynomial regression at wavelengths excluding the range 1.150–1.255 μm. Although the results could be affected by the fitting parameters and conditions (e.g., degree of polynomial regression and wavelength range), no significant absorption features appear regardless of the fitting parameters and conditions.

Since no clear absorption features are seen in the entire range of the observations, we provide upper limits to the abundances of some hydrated Cl-bearing salts based on the noise level of our observations. As an example, in the main text we show the results of our estimates of the noise level affecting the reflectance spectra near the center of the leading hemisphere (111 ± 25°W, 38 ± 16°S). This area is expected to contain the least amount of sulfur-bearing compounds originating from Io as well as the most abundant endogenic non-ice materials, because this area contains chaotic terrains and is located near the center of the leading hemisphere (Brown & Hand 2013; Fischer et al. 2015; 2017; Ligier et al. 2016). In addition, we also show the results of estimates for the western region in the leading hemisphere (160 ± 9°W, 3 ± 15°N) and for the region covering chaotic terrains near the center of the trailing hemisphere (256 ± 5°W, 15 ± 14°N). Fischer et al. (2015) suggested the highest abundances of anhydrous Cl-bearing salts in the former regions, while Ligier et al. (2016) reported the highest abundances of hydrated Cl-bearing salts in the latter regions.

The hydrated Cl-bearing salts proposed to exist on Europa have absorption features spanning the narrow wavelength range covering 1.167–1.226 μm (Figure 1). Figure 9(a) shows reflectance spectra at 111 ± 25°W, 38 ± 16°S of Europa at wavelengths of 1.150–1.255 μm, including that narrower wavelength range just mentioned. This figure shows that the reflectance spectra are curved owing to the broad absorption of $H_2O$ ice at ~1.3 μm. Figure 9(d) shows the difference between the reflectance spectra and the polynomial regression line of a cubic function obtained using the least-squares method. Based on the observed differences, the 1σ value of the noise level of the reflectance at 111 ± 25°W, 38 ± 16°S is $1.79 \times 10^{-3}$ in the wavelength range 1.150–1.255 μm. The 3σ value of the noise level at the observed area is greater than the absolute values of the differences between the reflectance spectra and the fitted spectra shown in Figure 8(b). Note that the 1σ value of the differences between the reflectance spectra and the fitted spectra of crystalline $H_2O$ ice plus $H_2SO_4$ ice are calculated as $2.46 \times 10^{-3}$ in wavelengths of 1.150–1.255 μm (Figure 8(b)). This value is comparable to the 1σ value of the noise level. In addition, Figures 9(b) and (c) show reflectance spectra at 160 ± 9°W, 3 ± 15°N and 256 ± 5°W, 15 ± 14°N. The 1σ values of the noise level of the reflectance at 160 ± 9°W, 3 ± 15°N and 256 ± 5°W, 15 ± 14°N are $2.39 \times 10^{-3}$ and $3.04 \times 10^{-3}$, respectively, as shown in Figures 9(e) and (f). We also calculated the 1σ values of the noise level of the reflectance in the other observed regions at wavelengths of 1.150–1.255 μm (see Tables B1 and C1 in the Appendix). The 1σ value of the noise level of the reflectance at 111 ± 25°W, 38 ± 16°S is close to the average value of the equivalent levels in the leading hemisphere (Appendix Tables B1 and C1).

To estimate the upper limits to the abundances of hydrated Cl-bearing salts, we used data available for the absorption depth of hydrated Cl-bearing salts (i.e., Hanley et al. 2014 for $Mg(ClO_4)_2\cdot6H_2O$, $Mg(ClO_3)_2\cdot6H_2O$, $MgCl_2\cdot4H_2O$, $MgCl_2\cdot6H_2O$, and $NaClO_4\cdot2H_2O$ at 80 K; Fox-Powell et al. 2019 for $NaCl\cdot2H_2O$ at 100 K). The absorption depths of hydrated salts were calculated based on the measured decrease in reflectance from the continuum levels (Hanley et al. 2014; Fox-Powell et al. 2019). The absorption centers of $Mg(ClO_4)_2\cdot6H_2O$, $Mg(ClO_3)_2\cdot6H_2O$, $MgCl_2\cdot4H_2O$, $MgCl_2\cdot6H_2O$, $NaClO_4\cdot2H_2O$, and $NaCl\cdot2H_2O$ are located at 1.170, 1.168, 1.202, 1.196, 1.166, and 1.226 μm, respectively (Hanley et al. 2014; Fox-Powell et al. 2019). Assuming a linear relationship between the absorption depth and abundance of materials as suggested by previous studies (e.g., Bauer et al. 2002), we calculated the amount of hydrated Cl-bearing salts with absorption depths comparable to the noise levels of the observations. Our calculations do not consider the widths of the potential absorption features of hydrated Cl-bearing salts with our wavelength resolution. If considering the widths of their absorptions with our wavelength resolution, the upper limits to the abundances of hydrated salts could be calculated more strictly than considering only their depths. However, these analyses would require the spectral modeling to calculate the width and shape of absorption features for different hydrated salts with various abundances and sizes. Thus, the present study calculated conservative upper limits to the abundances for each hydrated Cl-bearing salt based on comparing absorption depths with the noise levels. The data fisting of the observations using spectral modeling is future work of this study. In some cases, the noise level might be calculated by dividing the observational data by the fits. However, the absolute noise levels cannot be obtained in this method because the absorption depths of hydrated salts must also be divided by the absolute reflectance to compare with the noise. Thus, we compare the absolute absorption depths of hydrated Cl-bearing salts with the noise levels calculated by subtracting the regression line, rather than by dividing by the regression line.

Figure 10(a) shows the estimated absorption depth of hydrated Cl-bearing salts as a function of their abundance compared with the noise levels of the observations for the three regions (3σ noise levels). Figure 10(b) compares the calculated absorption of hydrated Cl-bearing salts with the upper limits to the abundances based on the residuals between the reflectance spectra and the regression curves of our observations. These figures show that the conservative upper limits to the abundances of most hydrated salts at 111 ± 25°W, 38 ± 16°S and 160 ± 9°W, 3 ± 15°N are ~17% at the 3σ noise level; whereas, the upper limit is 34% for $NaCl \cdot 2H_2O$. In particular, conservative upper limits to the abundances of $NaClO_4 \cdot 2H_2O$, $Mg(ClO_4)_2 \cdot 6H_2O$, $MgCl_2 \cdot 6H_2O$, and $MgCl_2 \cdot 4H_2O$ are estimated at 10% at the 3σ noise level: see Figure 10(a). Given that the noise level of the reflectance (111 ± 25°W, 38 ± 16°S and 160 ± 9°W, 3 ± 15°N) is comparable to the equivalent levels in the other observed areas in the leading hemisphere (Appendix Tables B1 and C1), the upper limits to the abundances of hydrated salts would be similar for the other observed areas as well. On the other hand, these figures also show that conservative upper limits to the abundances of most hydrated salts at 256 ± 5°W, 15 ± 14°N in the trailing hemisphere are ~11%–22% at the 3σ noise level, while the upper limit for $NaCl \cdot 2H_2O$ is 44%.

The estimated upper limits to the abundances of $NaClO_4 \cdot 2H_2O$, $Mg(ClO_4)_2 \cdot 6H_2O$, $Mg(ClO_3)_2 \cdot 6H_2O$, $MgCl_2 \cdot 6H_2O$, and $MgCl_2 \cdot 4H_2O$ are lower than the estimated abundance (~20%) of hydrated Cl-bearing salts in chaotic terrains in the leading hemisphere that is adopted to explain the low absorption of $H_2O$ ice (Ligier et al. 2016). In chaotic terrains in the trailing hemisphere, the estimated upper limits to the abundances of $NaClO_4 \cdot 2H_2O$, $Mg(ClO_4)_2 \cdot 6H_2O$, $Mg(ClO_3)_2 \cdot 6H_2O$, $MgCl_2 \cdot 6H_2O$, and $MgCl_2 \cdot 4H_2O$ are also lower than the estimated abundance (~30%) of hydrated Cl-bearing salts (Ligier et al. 2016). This implies that if the non-ice materials on Europa are Cl-bearing salts, the main component would include anhydrous Cl-bearing salts of NaCl and/or $NaClO_4$, or hydrated salt of $NaCl \cdot 2H_2O$, rather than other hydrated Cl-bearing salts such as $NaClO_4 \cdot 2H_2O$, $Mg(ClO_4)_2 \cdot 6H_2O$, $Mg(ClO_3)_2 \cdot 6H_2O$, $MgCl_2 \cdot 6H_2O$, or $MgCl_2 \cdot 4H_2O$. The present study uses the same library of laboratory data as Ligier et al. (2016) for hydrated Mg-bearing salts (Hanley et al. 2014), although the absorption depths of the reflectance spectra of hydrated salts can be affected by many factors (e.g., grain size and temperature; e.g., Hanley et al. 2014).

Although we cannot rule out the possibility that $NaCl \cdot 2H_2O$ is the major Cl-bearing salt on Europa based on the noise level of our observations, a previous experimental study has suggested that dehydration of $NaCl \cdot 2H_2O$ to NaCl would have proceeded effectively through irradiation by UV light and/or cycles of temperature changes on Europa (Thomas et al. 2017). Although the dehydration rate of $NaCl \cdot 2H_2O$ was not quantified in that previous study, the authors showed that dehydration occurred in a short time in laboratory experiments (e.g., irradiation by UV light for ~5 hours; Thomas et al. 2017). In addition, no previous observational data of Europa’s surface report any significant absorption features around 2.2 μm (e.g., McCord et al. 1999; Fischer et al. 2015), where $NaCl \cdot 2H_2O$ would exhibit an absorption feature (Fox-Powell et al. 2019). Thus, these previous observational studies might also support a lack of predominance of $NaCl \cdot 2H_2O$ as the major Cl-bearing salts, although the absorption feature near 2.2 μm may be due to impurities contained in the natural sample of the library data (e.g., gypsum; Fox-Powell et al. 2019). The above-mentioned experimental study also showed that dehydration of $NaCl \cdot 2H_2O$ proceeded more effectively than that of $MgCl_2 \cdot nH_2O$. Dehydrative conversion of $MgCl_2 \cdot 6H_2O$ to $MgCl_2 \cdot 4H_2O$ proceeded by means of UV-light irradiation; however, anhydrous $MgCl_2$ was not formed (Thomas et al. 2017). Combined with these previous experimental results (Thomas et al. 2017), our results showing no clear absorption features of $MgCl_2 \cdot nH_2O$ imply that if the non-ice materials on Europa are Cl-bearing salts, Na is more abundant than Mg in the salt composition. Note that the presence of sulfate salts suggested on Europa cannot be discussed based on our observational data, since most such salts have no distinct features within the wavelength range observed (e.g., Dalton et al. 2012).

Our conclusion suggesting a lack of Mg-containing salts on Europa supports the detection of $Na^+$ and $K^+$ with the non-detection of $Mg^{2+}$ in Europa's tenuous atmosphere (Brown 2001; Horst & Brown 2013) and a possible color center of NaCl in Europa's reflectance spectra at visible wavelengths (Trumbo et al. 2019); however, this might contradict a previous estimate by Ligier et al. (2016). The latter authors suggested that Europa's surface may contain $Mg(ClO_4)_2 \cdot 6H_2O$, $Mg(ClO_3)_2 \cdot 6H_2O$, and/or $MgCl_2 \cdot nH_2O$ based on observational data obtained using the VLT. In their spectral fitting model, addition of ~20% of hydrated Mg-bearing salts in the surface materials of the leading hemisphere of Europa improved their fits near the edge of the absorption of $H_2O$ ice (Ligier et al. 2016). This happened because the absorption tails of hydrated Mg-bearing salts decrease the reflectance near the edge of $H_2O$-ice absorption at 1.45 μm and 1.8 μm (Ligier et al. 2016). The proposed amount (~20%) of hydrated Mg-bearing salts is higher than our estimate of the upper limit to their abundance (8%–17% for the 3σ noise level). However, Ligier et al. (2016) found no absorption peaks of hydrated Mg-bearing salts. In addition, the absolute value and shape of the absorption of hydrated salts can be affected by many factors (e.g., grain size or temperature; e.g., Hanley et al. 2014). For instance, a reduction in the reflectance near the edge of the absorption of $H_2O$ ice could be caused by hydrated Mg-bearing salts with lower abundances if their grain size is larger. Since there are no experimental reports as to the optical constants of hydrated Mg-bearing salts at low temperatures, grain-size effects cannot be evaluated quantitatively through spectral fitting. Laboratory measurements of the optical constants of hydrated Mg-/Na-bearing salts would be useful to improve quantitative estimates of their abundances on Europa. These measurements of the optical constants would also enable a discussion of the presence of Cl-bearing salts on the surface, based on their absolute reflectance.

Alternatively, the apparent contradiction regarding the presence of hydrated Mg-bearing salts between our estimate and that of Ligier et al. (2016) may reflect a spatial heterogeneity of the composition of non-ice surface materials on Europa. The areas observed in the present study do not completely overlap with those of Ligier et al. (2016). The present study observed Europa at latitudes of 30°–60°S; whereas, Ligier et al. (2016) performed their observations mostly at latitudes of 40°N–40°S. Observational data at latitudes of 40°–50°S are sparse, and no data are available from Ligier et al. (2016) at latitudes of 50°–60°S.

If Na is more abundant than Mg in the non-ice materials on Europa, as suggested by the present study, this provides constraints on the water chemistry and water-rock interactions of the subsurface ocean. $Na^+$ tends to be the major cation when the mass ratio of seawater to seafloor rock is less than ~10 and/or the oceanic pH is circumneutral or alkaline (Zolotov & Kargel 2009; Tan et al. 2021); whereas, Mg can be the major cation when the W/R ratio is higher than ~10 or the pH is acidic (Zolotov & Kargel 2009). Assuming that non-ice materials reflect the water chemistry of the subsurface ocean, our conclusion supporting the presence of Na-rich salts suggests that the oceanic pH of Europa would be circumneutral or alkaline with a relatively low W/R ratio (<10). Zolotov & Kargel (2009) suggested that Europa's ocean can be acidic if sulfur in the ocean and Fe in seafloor rocks were oxidized owing to $H_2$ escape from the ocean. Therefore, our conclusion as to the presence of Na salts further implies that the seafloor rocks may not be fully oxidized and would sustain reducing power. This is important for geochemical sulfur cycles (Tan et al. 2021) and the habitability of potential chemoautotrophic life on Europa.

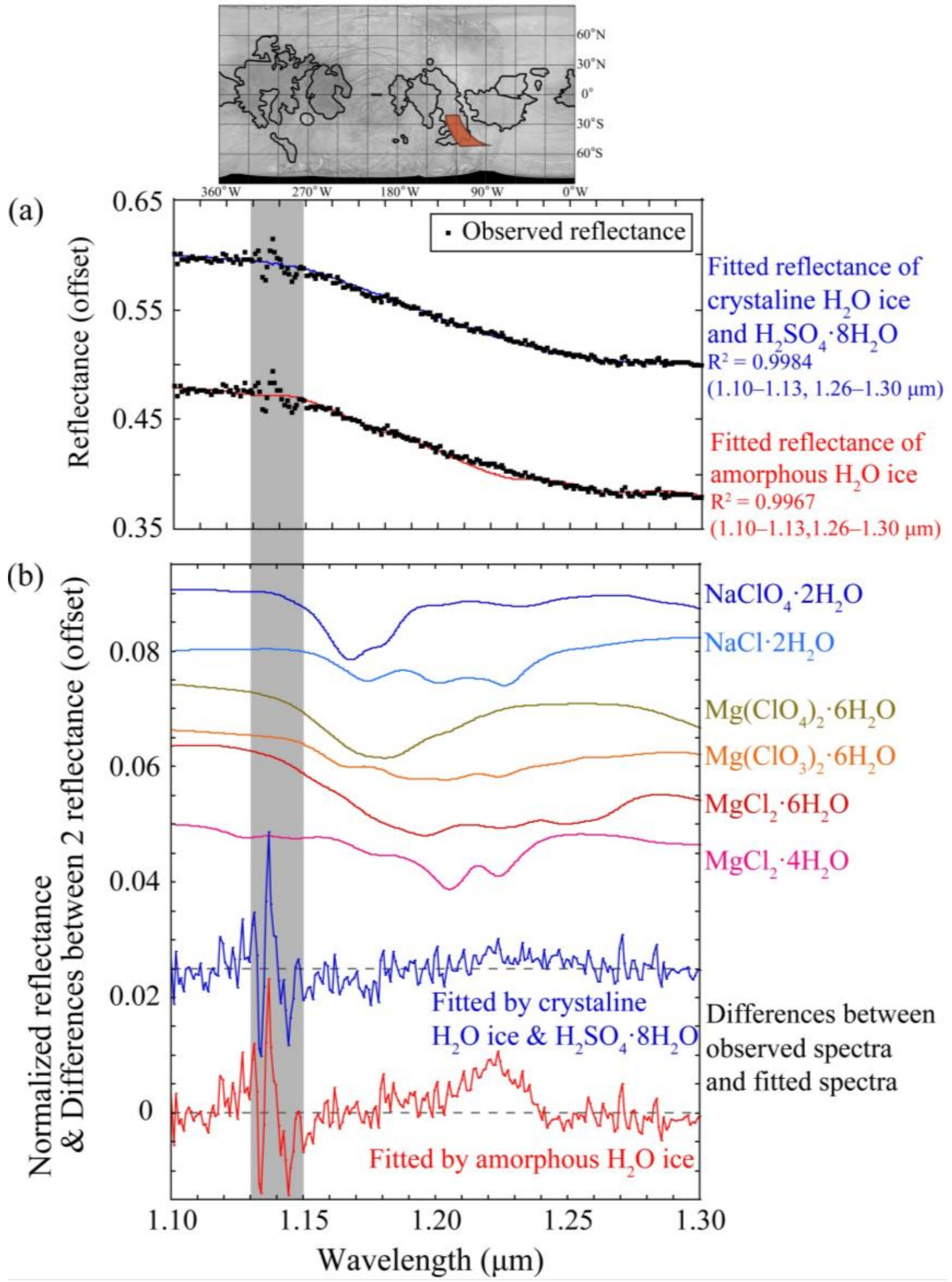

Figure 8. Analysis of the shape of the reflectance spectra at 111 ± 25°W, 38 ± 16°S at wavelengths of 1.100–1.300 μm. (a) Close-up view of the reflectance spectra fitted by the calculated $H_2O$-ice spectra. (b) Normalized spectra of hydrated Cl-bearing salts compared with differences between the observed and fitted $H_2O$-ice spectra as a function of wavelength. Reflectance spectra of $H_2O$ ice is calculated using the Hapke model (Hapke 1981; 1993; 2002) with the optical constants for amorphous $H_2O$ ice at 70–130 K (Mastrapa et al. 2008), crystalline $H_2O$ ice at 120 K (Grundy & Schmidt 1998), and $H_2SO_4 \cdot 8H_2O$ at 77 K (Carlson et al. 2005). The optical constants of $H_2O$ ice are supplemented with optical constants from Warren & Brandt (2008). The reflectance of $H_2O$ ice is fitted to the observed reflectance using the least-squares method at wavelengths of 1.100–1.130 and 1.255–1.300 μm. In the best fit with crystalline $H_2O$ ice

plus $H_2SO_4 \cdot 8H_2O$, the mixing ratios of $H_2O$ ice and $H_2SO_4 \cdot 8H_2O$ are calculated as 54% and 46%, respectively. Offsets have been added to the differences and the normalized spectra for clarification. The gray bar represents the wavelengths corresponding to telluric atmospheric absorption (1.13–1.15 μm). The normalized spectra of hydrated Cl-bearing salts are not scaled to their upper limits to the abundances.

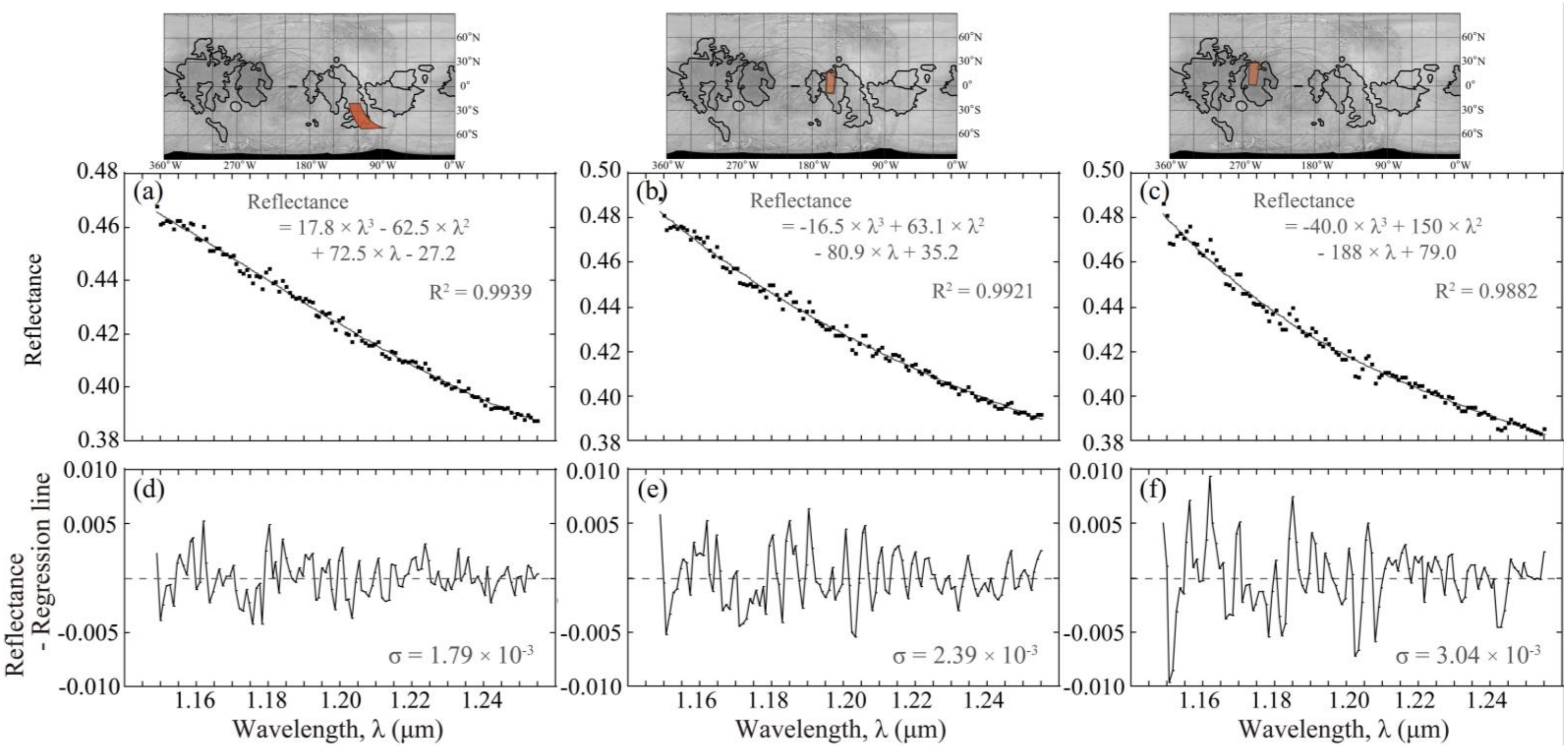

Figure 9. Results of our analysis of the noise level of the reflectance spectra at wavelengths of 1.150–1.255 μm. The observed areas on Europa are shown in the top insets. Close-up view of the reflectance spectra fitted by polynomial regression lines of a cubic function at (a) 111 ± 25°W, 38 ± 16°S, (b) 160 ± 9°W, 3 ± 15°N, and (c) 256 ± 5°W, 15 ± 14°N. Differences between the reflectance and a polynomial regression lines of the reflectance as a function of wavelength at (d) 111 ± 25°W, 38 ± 16°S, (e) 160 ± 9°W, 3 ± 15°N, and (f) 256 ± 5°W, 15 ± 14°N. The 1σ noise levels are calculated at $1.79 \times 10^{-3}$, $2.39 \times 10^{-3}$, and $3.04 \times 10^{-3}$, respectively.

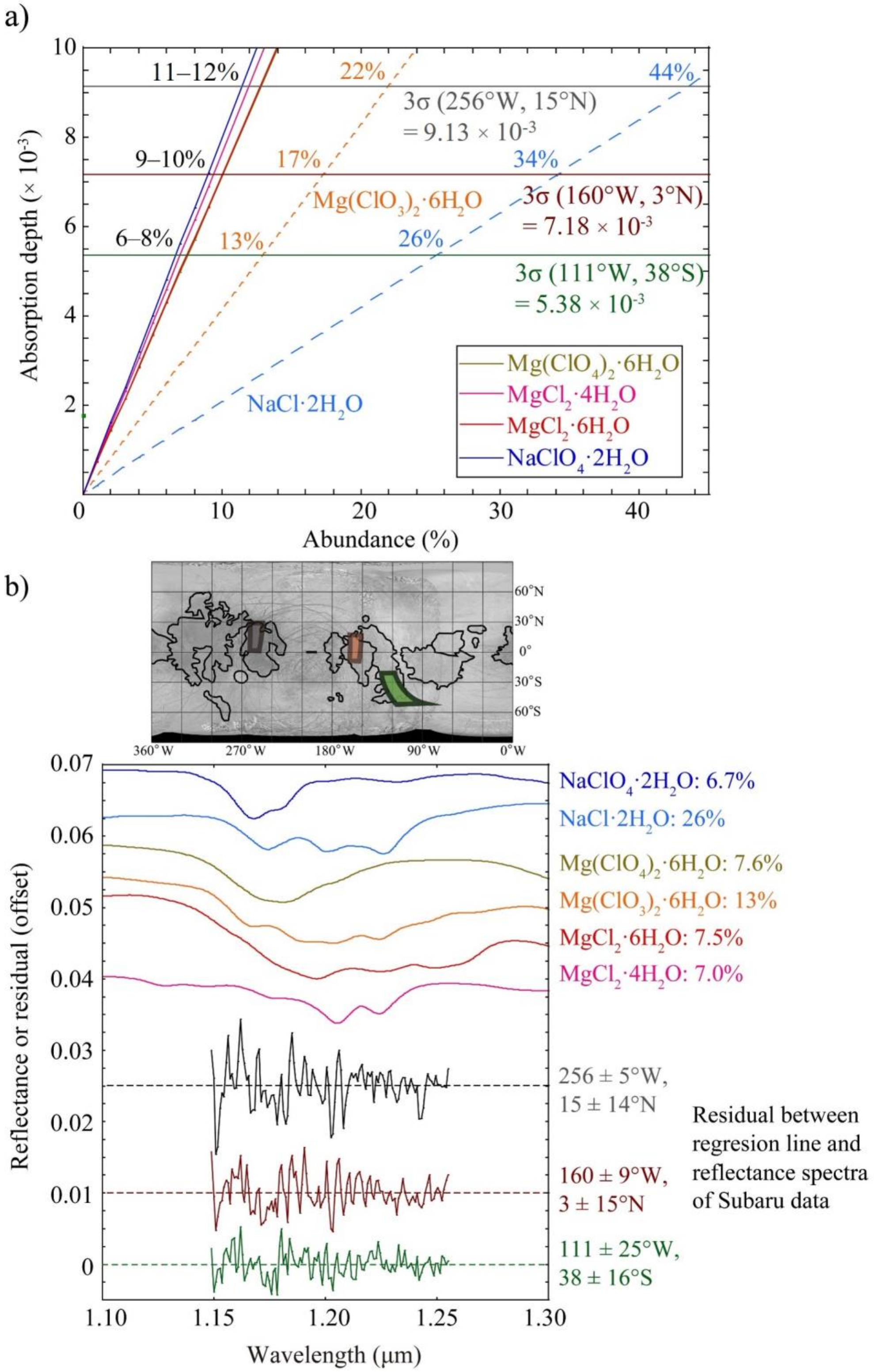

Figure 10. Estimated upper limits to the abundances of hydrated Cl-bearing salts in our observations. The observed regions on Europa are shown in the middle inset. (a) Estimated absorption depth of hydrated Cl-bearing

salts as a function of abundance compared with the noise levels of our observations. (b) Absorption of hydrated Cl-bearing salts for the estimated upper limits to the abundances at 111 ± 25°W, 38 ± 16°S compared with the residuals of the reflectance spectra fitted by the polynomial regression line at 1.150–1.255 μm (Figure 9(b)). Offsets have been added to the spectra of hydrated Cl-bearing salts and residuals for clarification.

## 5. CONCLUSIONS

Using the IRCS and AO system of the Subaru Telescope, we performed spatially resolved observations in the *zJH* band of Europa's surface. We obtained reflectance spectra in the wavelength range 1.0–1.8 μm with high wavelength resolution (δλ ~ 2 nm in the *zJH* band and δλ ~ 0.9 nm in the *H* band) and low noise level ($1\sigma = 1\text{–}2 \times 10^{-3}$). These wavelength resolutions and S/N ratios represent significant improvements with respect to previously available Galileo/NIMS data (δλ ~ 25 nm and S/N ~ 30; Carlson et al. 1992). In particular, the wavelength range 1.1–1.3 μm includes characteristic narrow absorption features due to hydrated Cl-bearing salts of $NaCl \cdot 2H_2O$, $NaClO_4 \cdot 2H_2O$, $MgCl_2 \cdot nH_2O$, $Mg(ClO_3)_2 \cdot 6H_2O$, and $Mg(ClO_4)_2 \cdot 6H_2O$; accordingly, our observations can constrain their presence on Europa. The conclusions of our observations are as follows:

- The reflectance spectra obtained of Europa's surface are smooth and featureless, except for absorption by $H_2O$ ice. We found no clear absorption features due to hydrated Cl-bearing salts around 1.2 μm.
- Based on the noise levels characteristic of our observations (see Section 4), the upper limits to the abundances of $NaClO_4 \cdot 2H_2O$, $MgCl_2 \cdot nH_2O$, and $Mg(ClO_4)_2 \cdot 6H_2O$ are estimated at ~10% at the $3\sigma$ noise level. The upper limit to the abundance is ~17% for $Mg(ClO_3)_2 \cdot 6H_2O$. These estimated values are lower than the abundances of Cl-bearing salts (> ~20% relative to $H_2O$ ice) required to explain the low $H_2O$ absorption found in previous observations (Ligier et al. 2016). This, in turn, suggests that anhydrous Cl-bearing salts (i.e., NaCl and $NaClO_4$) or hydrated $NaCl \cdot 2H_2O$ would be the major salts, if non-ice materials on Europa are Cl-bearing salts.

If surface salts reflect the subsurface ocean composition, our results suggest that $Na^+$ would be the major cation in Europa's subsurface ocean. $Na^+$

becomes dominant when the oceanic pH is circumneutral or alkaline and when sulfur in the ocean and Fe in the seafloor rocks are not oxidized (Zolotov & Kargel 2009). Thus, our conclusion implies that the seafloor rocks could sustain reducing power in water-rock reactions, which is important for the habitability and geochemistry of Europa.

## Acknowledgments

This research is based on data collected with the Subaru Telescope, which is operated by the National Astronomical Observatory of Japan (NAOJ). We are honored and grateful for the opportunity to observe the Universe from Maunakea, which has cultural, historical, and natural significance in Hawaii. Observational data are available from the Subaru–Mitaka–Okayama–Kiso Archive System (SMOKA; http://smoka.nao.ac.jp/) operated by NAOJ. Some of our data were obtained from SMOKA. We express our thanks to Y. Takagi and E. Mieda for their kind support during the operation of IRCS and AO188. Data analysis was in part carried out on the Multi-wavelength Data Analysis System operated by the Astronomy Data Center (ADC), NAOJ. This study was financially supported by Grants–in–aid for Scientific Research, KAKENHI JSPS (Grant Nos. JP17H06454 and JP17H06456).

## APPENDIX A. Signal counts from Europa's surface

Figure A1 shows the data of the extracted signal counts per second from the G2V stars at wavelengths of 0.95–1.85 μm. This figure indicates that there are relatively high levels of noise in the signal counts at wavelengths of 0.95–1.0 and 1.80–1.85 μm. This is due to the low efficiency of the grism spectrometer near the edges of the measured overall wavelength range. In addition, the spectra show high levels of signal scatter and low signal counts at wavelengths of 1.35–1.42 μm (Figure A1). This is because of the significant absorption by water vapor in the telluric atmosphere (e.g., Tokunaga et al. 2002).

Figure A2 shows the data of the extracted signal counts per second from Europa's surface at wavelengths of 0.95–1.85 μm from our observations. Since these reflectance spectra are spatially resolved, the fourteen spectra from the areas shown in Figure 5 are shown in Figure A2. All signal count spectra in Figure A2 contain relatively high levels of noise and low signal counts at 0.95–1.0 and 1.80–1.85 μm owing to the low efficiency of the grism spectrometer. The present study only uses signal counts from Europa at wavelengths of 1.0–1.8 μm to calculate the reflectance spectra of Europa's surface in Section 3.2 in the main text.

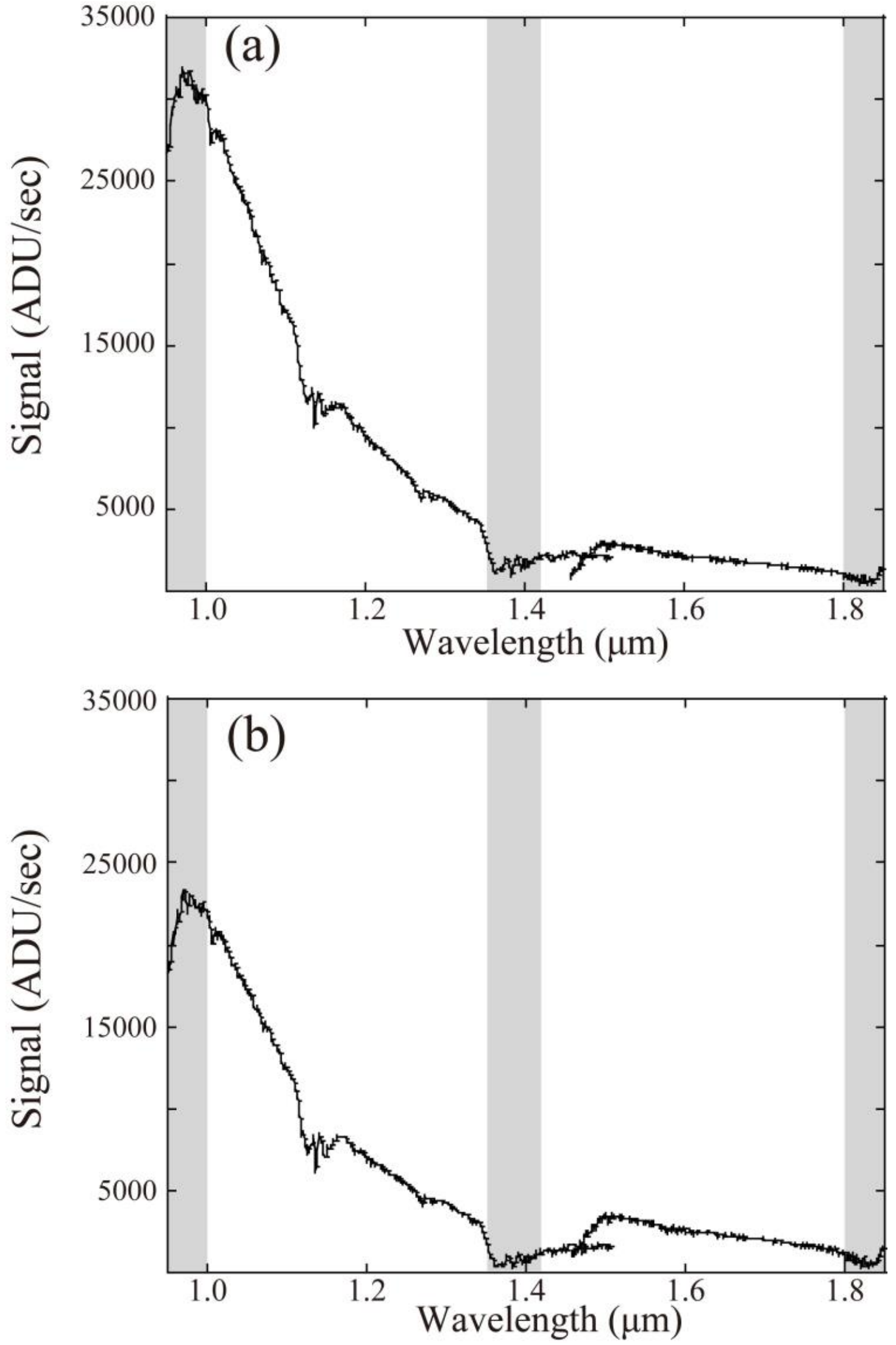

Figure A1. Extracted signal counts per second from the G2V stars (i.e., the 'telluric calibrators' in Table 1). Spectra in the *zJH* band (0.9–1.5 μm) and the *H* band (1.49–1.85 μm) were extracted from the spectral images of HD 154805 and HD 160257, respectively. (a) Data observed on 2019 May 17. (b) Data observed on 2019 May 18. In the gray-colored wavelength ranges, spectra contain high levels of noise due to telluric atmospheric

absorption (1.35–1.42 μm) and low efficiencies of the grism spectrometer (0.95–1.0 and 1.80–1.85 μm).

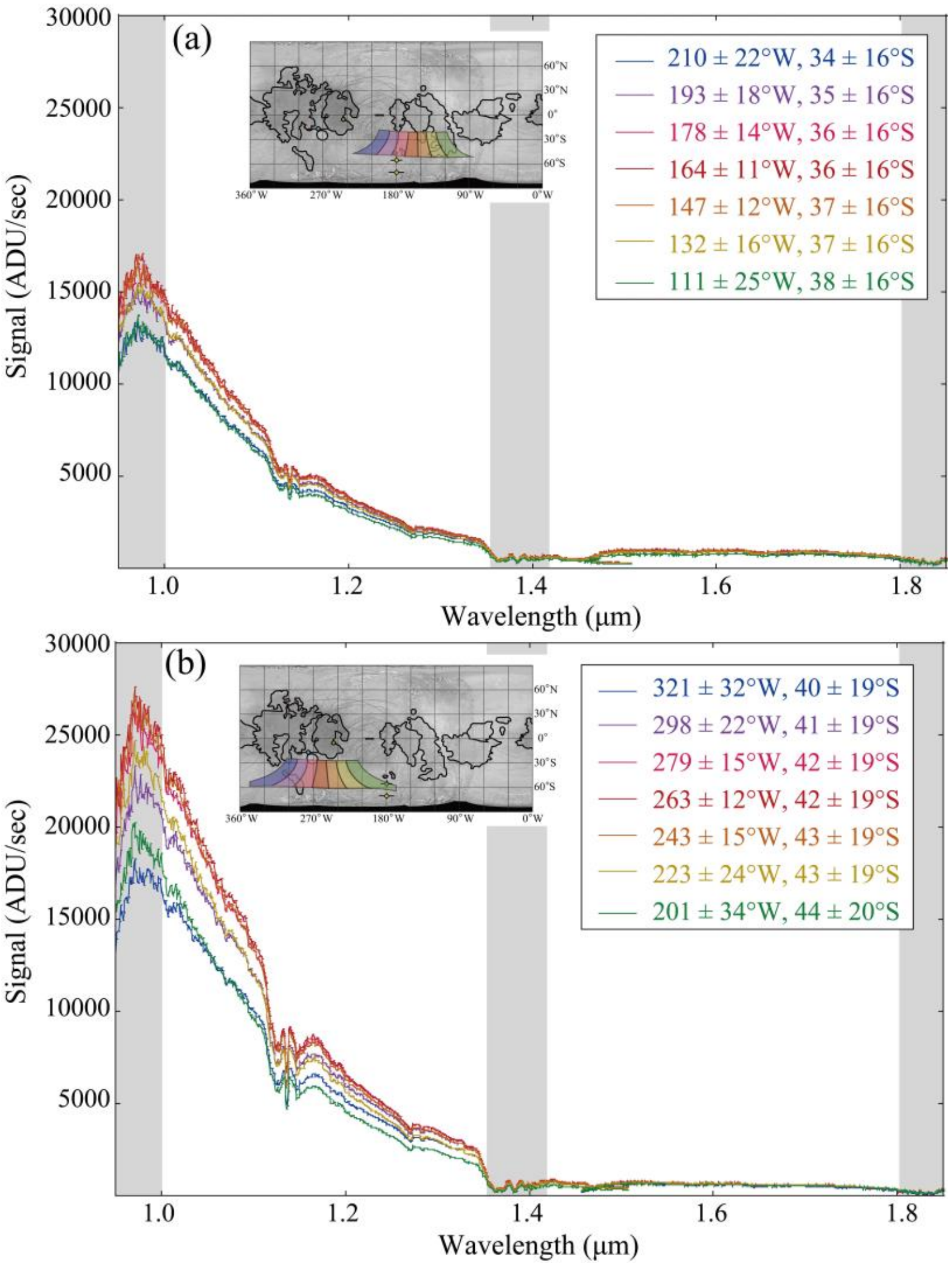

Figure A2. Extracted signal counts per second from Europa's surface. Each curve shows the spectra of one surface area (see the top insets for the locations). (a) Data for the southern regions in the leading hemisphere observed on 2019 May 17. (b) Data for the southern regions in the trailing hemisphere observed on 2019 May 18. In the gray-colored wavelength ranges, the spectra contain high levels of noise.

## APPENDIX B. Variations in the noise levels of the reflectance on Europa's surface

The present study calculates the noise levels of the reflectance spectra of all observed regions at wavelengths of 1.150–1.255 µm. The noise levels were calculated based on the differences between the reflectance and polynomial regression lines of the reflectance as a function of wavelength. Table B1 shows the noise levels at 1.50–1.94 × $10^{-3}$ in the observed regions in the leading hemisphere (Table B1). Thus, the 1σ value of the noise level of the reflectance of chaotic terrains (111 ± 25°W, 38 ± 16°S) is almost the average value of the levels in the leading hemisphere.

Table B1. Summary of 1σ values of the noise levels of the reflectance at wavelengths of 1.150–1.255 µm of the observed regions.

| Observed locations on Europa | 1σ reflectance values |
|---|---|
| 2019 May 17 (Leading hemisphere) | |
| 210 ± 22°W, 34 ± 16°S | 1.94 × $10^{-3}$ |
| 193 ± 18°W, 35 ± 16°S | 1.50 × $10^{-3}$ |
| 178 ± 14°W, 36 ± 16°S | 1.70 × $10^{-3}$ |
| 164 ± 11°W, 36 ± 16°S | 1.76 × $10^{-3}$ |
| 147 ± 12°W, 37 ± 16°S | 1.79 × $10^{-3}$ |
| 132 ± 16°W, 37 ± 16°S | 1.78 × $10^{-3}$ |
| 111 ± 25°W, 38 ± 16°S | 1.79 × $10^{-3}$ |
| 2019 May 18 (Trailing hemisphere) | |
| 321 ± 32°W, 40 ± 19°S | 2.23 × $10^{-3}$ |
| 298 ± 22°W, 41 ± 19°S | 1.80 × $10^{-3}$ |
| 279 ± 15°W, 42 ± 19°S | 1.85 × $10^{-3}$ |
| 263 ± 12°W, 42 ± 19°S | 2.25 × $10^{-3}$ |
| 243 ± 15°W, 43 ± 19°S | 2.12 × $10^{-3}$ |
| 223 ± 24°W, 43 ± 19°S | 2.17 × $10^{-3}$ |
| 201 ± 34°W, 44 ± 20°S | 2.58 × $10^{-3}$ |

## APPENDIX C. Observational data of the northern and equatorial areas on Europa

The present study observed the northern and equatorial regions of Europa in addition to the target areas discussed in the main text. Figure C1 shows the uncalibrated data of the extracted signal counts per second obtained for the other regions of Europa’s surface. Figure C2 shows calculated reflectance spectra for the other regions. The present study calculated the noise levels of the reflectance spectra of the other observed regions at wavelengths of 1.150–1.255 μm (Table C1).

The original signal data and calculated reflectance spectra for the northern and equatorial areas are shown in Figures C1 and C2, respectively. Figure C2 shows that the slopes at 1.0–1.3 μm become steep in regions close to the north pole. The slopes in the northern and equatorial regions are steeper than those derived from Galileo/NIMS spectra (McCord et al. 1999). This trend of the slopes does not appear in the east-west direction. A possible cause that could have resulted in steep slopes in the northern and equatorial areas is the dependence of AO performance on wavelength. The PSF of standard G2V stars would be larger, closer to ~1.0 μm due to the wavelength dependence of the Strehl ratios of the Subaru AO system (Minowa et al. 2010). This would cause lower flux levels of the standard G2V stars at shorter wavelengths close to 1.0 μm. This may have caused high reflectance at short wavelengths, and this could cause steep slopes in Europa’s reflectance spectra; however, this might not explain the latitude dependence of the effect. An alternative possibility is that the steep slopes in the northern and equatorial regions may reflect a difference in the surface physical properties (e.g., grain size, abundance) of ice. Our observed regions are not exactly the same regions as those of Galileo/NIMS data (McCord et al. 1999). Regional differences in the surface properties might result in differences in the reflectance slope. In addition, a similar regional difference in the reflectance slope at 1.1–1.3 μm was recently reported (Ligier et al. 2020). These authors suggest that the regional difference is consistent with the Galileo/NIMS data. Although their wavelength range 1.1—1.35 μm is not exactly the same as our wavelength range 1.0—1.8 μm, the differences in the reflectance slope may be partly explained by such regional differences. However, such a difference in the reflectance slope has not been reported clearly; accordingly, the difference in the reflectance slope may have been caused by other reasons. In contrast, the slopes at 1.0–1.3 μm in the southern regions are similar to those obtained from the Galileo/NIMS data (McCord et al. 1999). Thus, the present study uses the southern regions as target areas for detailed analysis (see Section 4 in the main text). At this stage, we cannot determine the reasons underlying the difference in the reflectance slope in the northern and equatorial regions. Although we did not use the data for the northern and equatorial regions, there are no clear absorption features at 1.2 μm in the spectra covering these areas (Figure C2). As such, our conclusion that the abundances of hydrated Cl-bearing salts are very low (see Section 4 in the main text) does not change upon consideration of the northern and equatorial regions.

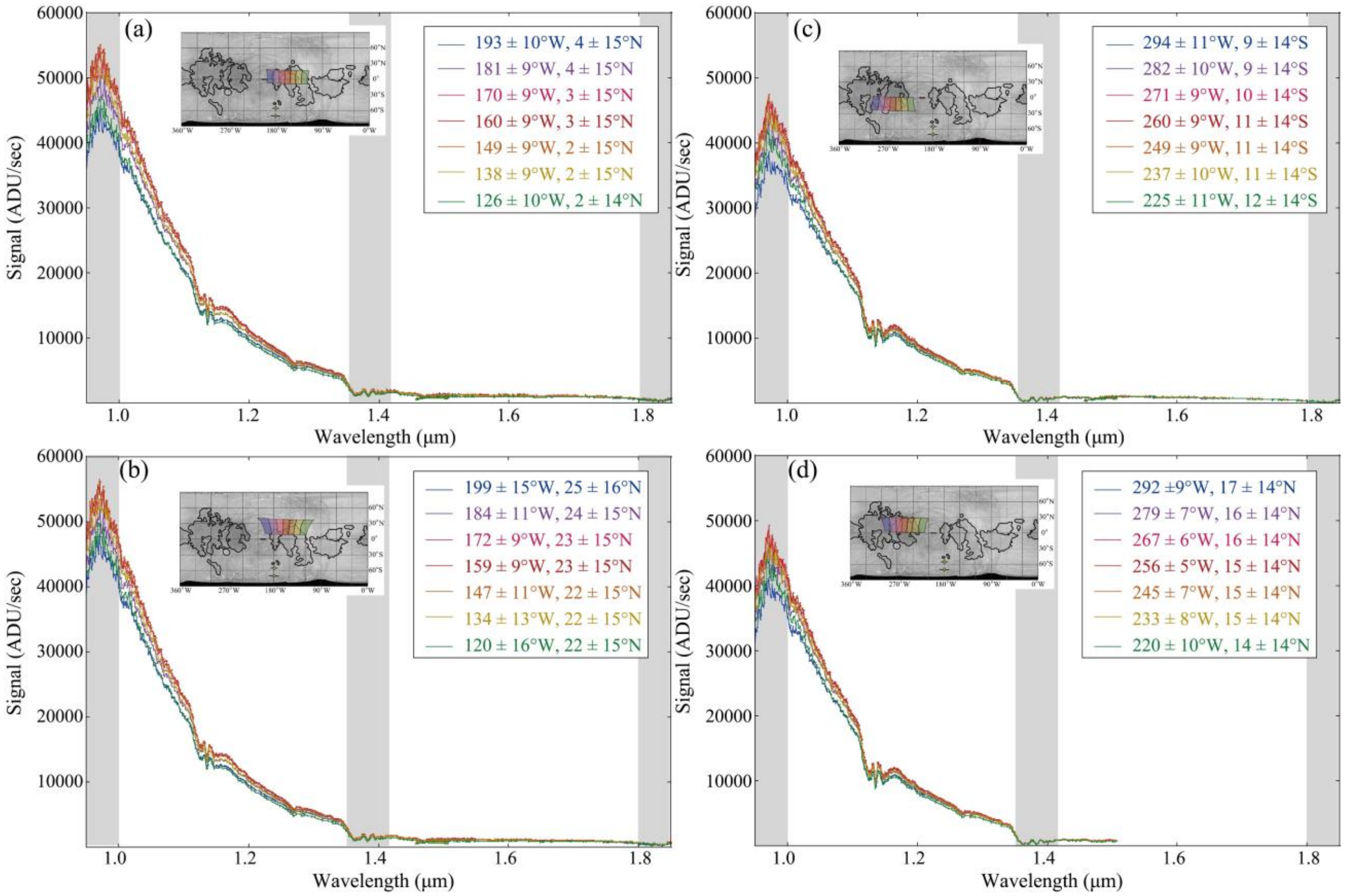

Figure C1. Extracted signal counts per second from the northern and equatorial regions of Europa’s surface. Each curve shows spectra for a given surface area (see the top insets for the corresponding locations). Data for the (a) equatorial and (b) northern regions in the leading hemisphere were taken on 2019 May 17. Data for the (c) equatorial and (d)

northern regions in the trailing hemisphere were taken on 2019 May 18. In the gray-colored wavelength ranges, the spectra are affected by high levels of noise.

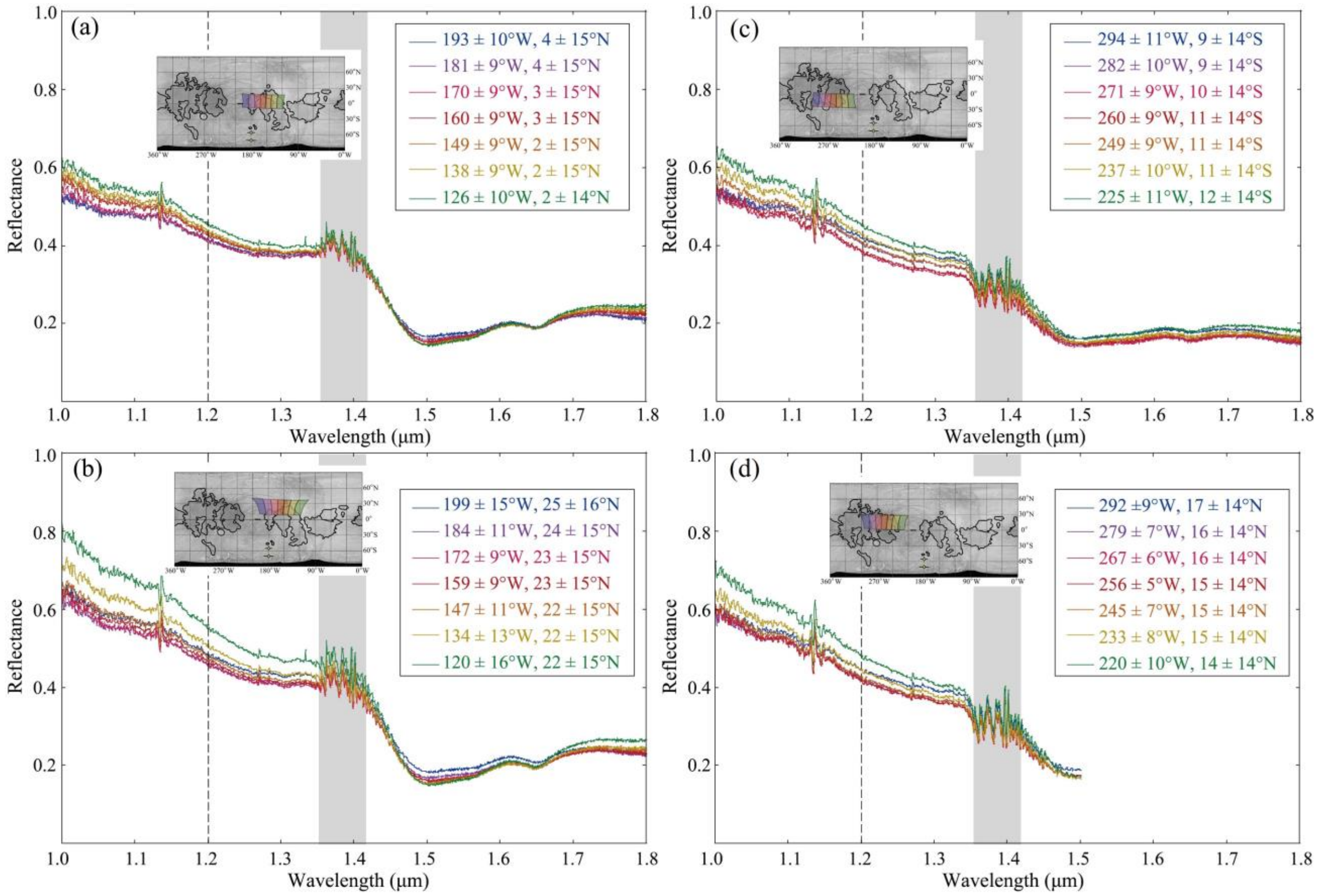

Figure C2. Connected reflectance spectra of the northern and equatorial regions on Europa at wavelengths of 1.0–1.8 μm. The observed area on Europa is shown in the top-left panel. Data for the (a) equatorial and (b) northern regions in the leading hemisphere were taken on 2019 May 17. Data for the (c) equatorial and (d) northern regions in the trailing hemisphere were taken on 2019 May 18. Gray bars represent the wavelengths corresponding to telluric atmospheric absorption features (1.35–1.42 μm). Vertical dashed lines indicate at a wavelength of 1.2 μm. In the reflectance spectra, we observe no significant absorption features around 1.2 μm, where hydrated Cl-bearing salts exhibit their characteristic absorption features. The slopes at 1.0–1.3 μm become steeper in regions close to the north pole.

Table C1. Summary of 1σ values of the noise levels of the reflectance at wavelengths of 1.150–1.255 μm of the northern and equatorial regions.

| Observed locations on Europa | 1σ reflectance values |
|---|---|
| 2019 May 17 (Leading hemisphere) | |
| 193 ± 10°W, 4 ± 15°N | $2.25 \times 10^{-3}$ |
| 181 ± 9°W, 4 ± 15°N | $2.04 \times 10^{-3}$ |
| 170 ± 9°W, 3 ± 15°N | $2.28 \times 10^{-3}$ |
| 160 ± 9°W, 3 ± 15°N | $2.39 \times 10^{-3}$ |
| 149 ± 9°W, 2 ± 15°N | $2.43 \times 10^{-3}$ |
| 138 ± 9°W, 2 ± 15°N | $2.31 \times 10^{-3}$ |
| 126 ± 10°W, 2 ± 14°N | $2.28 \times 10^{-3}$ |
| 199 ± 15°W, 25 ± 16°N | $2.82 \times 10^{-3}$ |
| 184 ± 11°W, 24 ± 15°N | $2.57 \times 10^{-3}$ |
| 172 ± 9°W, 23 ± 15°N | $2.78 \times 10^{-3}$ |
| 152 ± 9°W, 23 ± 15°N | $2.87 \times 10^{-3}$ |
| 147 ± 11°W, 22 ± 15°N | $3.00 \times 10^{-3}$ |
| 134 ± 13°W, 22 ± 15°N | $2.99 \times 10^{-3}$ |
| 120 ± 16°W, 22 ± 15°N | $3.03 \times 10^{-3}$ |
| 2019 May 18 (Trailing hemisphere) | |
| 294 ± 11°W, 9 ± 14°S | $2.68 \times 10^{-3}$ |
| 282 ± 10°W, 9 ± 14°S | $2.43 \times 10^{-3}$ |
| 271 ± 9°W, 10 ± 14°S | $2.31 \times 10^{-3}$ |
| 260 ± 9°W, 11 ± 14°S | $2.62 \times 10^{-3}$ |
| 249 ± 9°W, 11 ± 14°S | $2.77 \times 10^{-3}$ |
| 237 ± 10°W, 11 ± 14°S | $3.04 \times 10^{-3}$ |
| 225 ± 11°W, 12 ± 14°S | $2.65 \times 10^{-3}$ |
| 292 ± 9°W, 17 ± 14°N | $2.97 \times 10^{-3}$ |
| 279 ± 7°W, 16 ± 14°N | $2.82 \times 10^{-3}$ |
| 267 ± 6°W, 16 ± 14°N | $2.86 \times 10^{-3}$ |
| 256 ± 5°W, 15 ± 14°N | $3.04 \times 10^{-3}$ |
| 245 ± 7°W, 15 ± 14°N | $3.18 \times 10^{-3}$ |
| 233 ± 8°W, 15 ± 14°N | $3.15 \times 10^{-3}$ |
| 220 ± 10°W, 14 ± 14°N | $3.16 \times 10^{-3}$ |

References

Baba, H. et al. 2002, in ASP Conf. Ser. 281, ADASS XI, ed. D. A. Bohlender et al. (San Francisco, CA: ASP), 298

Bauer, J. M., Roush, T. L., Geballe, T. R., et al. 2002, Icar, 158, 178

Brown, M. E. 2001, Icar, 151, 190

Brown, M. E., & Hand, K. P. 2013, AJ, 145, 110

Carlson, R. W., Anderson, M. S., Johnson, R. E., et al. 2002, Icar, 157, 456

Carlson, R. W., Anderson, M.S., Mehlman, R., Johnson, R. E. 2005, Icar, 177, 461

Carlson, R. W., Calvin, W. M., Dalton, J. B., et al. 2009, in Europa, ed. R. T. Pappalardo et al. (Tucson, AZ: Univ. Arizona Press), 283

Carlson, R. W., Johnson, R. E., & Anderson, M. S. 1999, Sci, 286, 97

Carlson, R. W., Weissman, P. R., Smythe, W. D., et al. 1992, SSRv, 60, 457

Dalton, J. B., Shirley, J. H., & Kamp, L. W. 2012, JGR, 117, E03003

Dalton, J. B., Cassidy, T., Paranicas, C., et al. 2013, P&SS, 77, 45

Dalton, J. B. 2007, GeoRL, 34, L21205

Doggett, T., Greeley, R., & Figueredo, P. H. 2009, in Europa, ed. R. T. Pappalardo et al. (Tucson, AZ: Univ. Arizona Press), 137

Fischer, P. D., Brown, M. E., & Hand, K. P. 2015, AJ, 150, 164

Fox‐Powell, M. G., Osinski, G. R., Applin, D., et al. 2019, GeoRL, 46, 5759

Greenberg, R., Hoppa, G.V., Tufts, B.R., et al. 1999, Icar, 141, 263

Grundy, W. M., & Schmidt, B. 1998, JGR, 103, 25809

Hand, K. P., & Carlson, R. W. 2015, GeoRL, 42, 3174

Hanley, J., Dalton, J. B., Chevrier, V. F., et al. 2014, JGRE, 119, 2370

Hapke, B. 1981, JGR, 86, 3039

Hapke, B. 1993, Theory of Reflectance and Emittance Spectroscopy (Cambridge: Cambridge Univ. Press)

Hapke, B. 2002, Icar, 157, 523

Hayano, Y., Takami, H., Oya, S., et al. 2010, Proc. SPIE, 7736, 21

Hayano, Y., Takami, H., Guyon, O., et al. 2008, Proc. SPIE, 7015, 25

Hibbitts, C. A., Stockstill-Cahill, K., Wing, B., et al., Icar, 326, 37

Hog, E., Fabricius, C., Makarov, V. V., et al. 2000, A&A, 355, 27
Horst, S. M., & Brown, M. E. 2013, ApJ, 764, L28
Hussmann, H., & Spohn, T. 2004, Icar, 171, 391
Kargel, J. S., Kaye, J. Z., Head, J. W., et al. 2000, Icar, 148, 226
Kivelson, M.G., Khurana, K.K., Russell, C.T., et al. 2000, Sci, 289, 1340
Kobayashi, N., Tokunaga, A. T., Terada, H., et al. 2000, Proc. SPIE, 4008, 1056
Ligier, N., Poulet, F., Carter, J., et al. 2016, AJ, 151, 163
Ligier, N., Carter, J., Poulet, F., et al. 2020, LPSC 51, 1964
Mastrapa, R. M., Bernstein, M. P., Sandford, S. A., et al. 2008, Icar, 197, 307
McCord, T. B., Hansen, G. B., Matson, D. L., et al. 1999, JGRE, 104, 11827
McCord, T. B., Hansen, G. B., Clark, R. N., et al. 1998, JGR, 103, 8603
Minowa, Y., Hayano, Y., Oya, S., et al. 2010, Proc. SPIE, 7736, 77363N
Moore, W. B., & Hussmann, H. 2009, in Europa, ed. R. T. Pappalardo et al. (Tucson, AZ: Univ. Arizona Press), 369
O'Brien, D. P., Geissler, P., & Greenberg, P. 2002, Icar, 156, 152
Ockman, N. 1958, AdPhy, 7, 199
Seitz, F. 1946, RvMP, 18, 384
Tan, S., Sekine, Y., Shibuya, T., et al. 2021, Icar, 357, 114222
Thomas, E. C., Hodyss, R., Vu, T. H., et al. 2017, ESC, 1, 14
Tody, D. 1986, Proc. SPIE, 0627, 733
Tody, D. 1993, ASPC 52, Astronomical Data Analysis Software and Systems II, ed. R.J. Hanisch, R.J.V. Brissenden, & J. Barnes (San Francisco, CA: ASP), 173
Tokunaga, A. T., Kobayashi, N., Bell, J., et al. 1998, Proc. SPIE, 3354, 512
Tokunaga, A. T., Simons, D. A., & Vacca, W. D. 2002, PASP, 114, 180
Trumbo, S. K., Brown, M. E., & Hand, K. P. 2019, SciA, 5, eaaw7123
Trumbo, S. K., Brown, M. E., & Hand, K. P. 2020, AJ, 160, 282
Warren, S. G., & Brandt, R. E. 2008, JGR, 113, D14220
Zolotov, M. Y., & Kargel, J. S. 2009, in Europa, ed. R. T. Pappalardo et al. (Tucson, AZ: Univ. Arizona Press), 431